\documentclass[11pt,fleqn]{article}
\usepackage[utf8]{inputenc}
\usepackage{graphicx}
\usepackage{float}
\usepackage{amssymb,amsmath,amsfonts, mathtools}
\usepackage{enumerate}
\usepackage{xcolor}
\usepackage{dsfont}
\usepackage{comment}
\usepackage{amsthm} 
\usepackage{amsfonts,euscript,amssymb,stmaryrd,braket}

\usepackage{verbatim}
\usepackage{amsfonts,euscript,amssymb,stmaryrd,braket}
\usepackage{graphics,tikz}
\usetikzlibrary{arrows,decorations.markings,patterns}
\usepackage{caption}
\usepackage{subcaption}
\usepackage{slashed}
\definecolor{hyperref}{RGB}{026,028,185}
\usepackage[bookmarks=true,colorlinks=true,linkcolor=hyperref,citecolor=hyperref,urlcolor=hyperref,bookmarksnumbered]{hyperref}
 \usepackage{booktabs}
 \usepackage{multirow}
\usepackage{tabularx}
\usepackage{longtable}
\usepackage{bbold,bbding}
\usepackage{amsthm}

\definecolor{hyperref}{RGB}{026,028,185}
\usepackage[bookmarks=true,colorlinks=true,linkcolor=hyperref,citecolor=hyperref,urlcolor=hyperref,bookmarksnumbered]{hyperref}
\usepackage{footmisc}

\def\clock{{\count0=\time
		\divide\count0 60
		\ifnum\count0<10 0\fi\the\count0
		\multiply\count0 -60 \advance\count0 \time
		:\ifnum\count0<10 0\fi \the\count0
}}
\newcommand{\timestamp}{{\small\vbox{\hbox{\tt\jobname.tex}
			\hbox{\the\day/\the\month/\the\year, \clock}}}}

\newcommand{\ba}{\begin{eqnarray}}
\newcommand{\ea}{\end{eqnarray}}

\renewcommand{\deg}{\text{deg}\,}

\newcommand{\be}{\begin{equation}}
\newcommand{\ee}{\end{equation}}

\makeatletter
\let\old@startsection=\@startsection
\let\oldl@section=\l@section
\renewcommand{\@startsection}[6]{\old@startsection{#1}{#2}{#3}{#4}{#5}{#6\mathversion{bold}}}
\renewcommand{\l@section}[2]{\oldl@section{\mathversion{bold}#1}{#2}}
\makeatother

\numberwithin{equation}{section}
\usepackage{color}

\def\no{\nonumber}
\def\a{\alpha}
\def\b{\beta}

\def \adss {$AdS_5 \times S^5$\ }

\newcommand{\beq}{\begin{equation}}
\newcommand{\eeq}{\end{equation}}

\renewcommand{\a}{\alpha}
\renewcommand{\b}{\beta}

\renewcommand{\S}{\Sigma}

\newcommand{\Tr}{\textup{Tr}}

\newcommand{\bal}{\begin{equation}\begin{aligned}}
\newcommand{\eal}{\end{aligned} \end{equation}}

\begin{document}
	\renewcommand{\thefootnote}{\arabic{footnote}}
	
	\overfullrule=0pt
	\parskip=2pt
	\parindent=12pt
	\headheight=0in \headsep=0in \topmargin=0in \oddsidemargin=0in
	
	\vspace{ -3cm} \thispagestyle{empty} \vspace{-1cm}
	\begin{flushright} 
		\footnotesize
		HU-EP-22/36-RTG  
	\end{flushright}%

	\begin{center}
		\vspace{1.2cm}
		{\Large\bf \mathversion{bold}
			 Holography on the lattice:\\ the string worldsheet perspective
 %solution
		%String worldsheet for the null cusp in lattice perturbation theory
		}
		
		\author{Gabriel~Bliard\thanks{XYZ} \and DEF\thanks{UVW} \and GHI\thanks{XYZ}}
		\vspace{0.8cm} {
			Gabriel~Bliard$^{~a,}$\footnote{{\tt $\{$gabriel.bliard,ilaria.costa$\}$@\,physik.hu-berlin.de}}, Ilaria~Costa$^{a,1}$, 
			Valentina~Forini$^{b,a}$\footnote{\tt valentina.forini@city.ac.uk}}
		\vskip  0.5cm
		
		\small
		{\em
			$^{a}$	Institut f\"ur Physik, Humboldt-Universit\"at zu Berlin, IRIS Adlershof, \\Zum Gro\ss en Windkanal 2, 12489 Berlin, Germany  
			\vskip 0.05cm
			$^{b}$      Department of Mathematics, City, University of London\\
			Northampton Square, EC1V 0HB London, United Kingdom }
		\normalsize
		
	\end{center}

	\vspace{0.3cm}
	\begin{abstract} 
		We review the study, on the lattice, of the Green-Schwarz gauge-fixed string action describing worldsheet fluctuations about the minimal surface holographically dual to the null cusp Wilson loop, useful to evaluate the cusp anomaly of $\mathcal{N}=4$ super Yang-Mills (sYM).  We comment on discretization, numerical explorations and challenges for the nonperturbative study of this benchmark model of gauge-fixed worldsheet actions.  
	\end{abstract}
	
	\newpage
	
   \setcounter{footnote}{0}
   
	%%%%%%%%%%%%%%%%%%%%%%%%%%%%%%%%%%%%%%%%%%%%%
	%%%%%%%%%%%%%%%%%%%%%%%%%%%%%%%%%%%%%%%%%%%%%
	\tableofcontents

\section{Overview}

The string side of the AdS/CFT correspondence, in its best studied case, is a Green-Schwarz (GS) superstring nonlinear sigma-model with target space AdS$_5\times$ S$^5$.  
Formal arguments exist for its UV finiteness at all orders~\cite{MT1998,Roiban:2007jf}, relying on its abundance in global and local symmetries. 
For its quantization one usually proceeds in a semiclassical fashion, expanding around classical solutions in a gauge-fixed setup. In relevant cases, the model turns out to be a relatively complicated, non polynomial action with non-trivial fermionic interactions. UV finiteness is non manifest, but adopting dimensional regularization it has been verified up to two loop order, the current limit of sigma-model perturbation theory. Once the string action is gauge-fixed and expanded around a chosen classical solution, the model becomes effectively a (non-trivial) $d=2$ field theory of worldsheet excitations. Given its expected UV finiteness, the question of its nonperturbative definition via field theory methods which exploit a lattice discretization of the worldsheet is a legitimate one.  At least in principle, the lower dimensionality  and the absence of $d=2$ supersymmetry constitute advantages with respect to the four-dimensional  lattice gauge field theory approach to holography (see related articles of this special issue).  

A perfect framework for these investigations is the effective worldsheet action for excitations around the string configuration dual to the cusped lightlike  Wilson loop of~\cite{Gubser:2002tv}
-- reviewed below in the  setup of~\cite{Giombi:2009gd}. The string partition function weighted by this action defines the so-called ``cusp anomalous dimension'' of $\mathcal{N}=4$ super Yang-Mills, which is accessible at all orders through the assumption of integrability~\cite{Beisert:2006ez}. The rich physics underlying this setup and the remarkable availability of finite-coupling predictions explain the “benchmark” role of this action in the pioneering studies~\cite{McKeown:2013vpa, Bianchi:2016cyv,Forini:2016sot,Forini:2017mpu,Bianchi:2019ygz, Bliard:2022kne}.  The setup proposed in~\cite{Bianchi:2016cyv,Bianchi:2019ygz} has been used to perform  Monte Carlo simulations in order to measure both the cusp anomaly and bosonic and fermionic correlators.  In these explorations a number of theoretical and numerical aspects were addressed and partially solved: for example, a complex phase affecting the fermionic  sector in~\cite{Bianchi:2016cyv} was eliminated in~\cite{Bianchi:2019ygz}, numerical instabilities were cured in a large region of parameter space adopting procedures (twisted-mass reweighting) standard in lattice QCD; also, this setup has been used to test an interesting stochastic algorithm for the estimation of the trace of implicit matrices~\cite{Forini:2021keh}, which was so far used only in the machine learning community and may prove useful in other high-energy-physics areas. However, the Wilson-like fermionic discretization adopted in~\cite{Bianchi:2016cyv,Bianchi:2019ygz} breaks part of the global symmetry of the model. The resulting divergences observed e.g. in simulations for fermionic correlators~\cite{Bianchi:2019ygz} urged for further investigation. More recently,  a new discretization has been worked out for the same model, which  is invariant under the full group of global symmetries - and therefore it is the best premise for investigations~\cite{Bliard:2022kne}. As we review below, however, a lattice perturbation theory analysis of one loop renormalizability reveals that the situation is much more complicated than in dimensional regularization, and it is related to the presence of power UV divergences. In order to remove them at one loop, it is necessary to introduce two extra parameters in the action, which need to be either fine-tuned at tree level or renormalized at one loop. 
These do not seem to have any deep meaning, besides the fact that they make the bare propagators particularly simple.

That the one-loop fine-tuning of~\cite{Bliard:2022kne} is enough to make all physical observables finite at all orders in perturbation theory seems unlikely. It would be still important to understand whether the number of parameters needed to achieve finiteness of all correlators - via fine-tuning or renormalization  - is finite. If not, the discretized model has no predictivity. A complete one-loop analysis of the divergences of $n$-point functions, e.g. on the general lines of~\cite{Gasser:1983yg}  and relying on the technology developed in~\cite{Bliard:2022kne}, would be necessary here.  This calculation would be easily generalisable to a number of observables and backgrounds.  Obviously, a positive answer to this question would not be a definite statement on the nonperturbative renormalizability of the model, but a strong indication of the action of symmetries.  In addition, notice that  the divergences requiring the introduction and tuning of extra-parameters in~\cite{Bliard:2022kne} are of linear type and very specific of the
lattice discretization.  One may try to find a general mechanism that prevents linear divergences in the first place, trying to exploit a spurionic symmetry involving 
the replacement $m \to -m$, the reflection of both worldsheet coordinates and an $SO(5)$ rotation, which is enjoyed by the continuous action. Preliminary explorations seem to indicate that it is not completely trivial to preserve this symmetry on the lattice while avoiding the doubling problem. 
A crucial point is that the symmetries preserved in~\cite{Bliard:2022kne} are only a small subset of the set of symmetries initially present of the AdS$_5\times S^5$ superstring sigma-model:  target space supersymmetry -- the $PSU(2,2|4)$ supergroup --  is broken and local symmetries (diffeomorphism and fermionic kappa-symmetry) are all fixed. This holds for the large majority of the studied string configurations dual to interesting gauge theory observables, but not much has been said, in the continuum, on the ``remnants'' of these symmetries in terms of a nonlinear realization, or of ``BRST''-like mechanism (for the local ones) and related Ward identities. An explicit investigation of this kind, in particular for the fermionic $\kappa$-symmetry -- which plays a central role in the formal argument for UV finiteness of the GS string  -- appears important in the continuum in the first place (in a related setup, some discussion on $\kappa$-symmetry ghost contributions appears in~\cite{Roiban:2007dq}).
Then, one would have to search for a discretization able to preserve (at least a subgroup of) such remnant symmetries~\cite{Capitani:2002mp}, and study the corresponding continuum limit. %~\cite{Capitani:2002mp}. 

We conclude by remarking that, in many known cases, symmetries play a crucial role in proving renormalizability of theories that seem to be non-renormalizable on the basis of power counting alone (as it happens for the discretized cusp model of~\cite{Bliard:2022kne}, as we see below). This is what happens for the well-known 2d $O(N)$ non-linear sigma-model both in dimensional regularization and on the lattice~\cite{Brezin:1975sq}, and it would be pedagogical -- to go closer to the superstring worldsheet setup --  to conduct numerical  explorations in the setup obtained adding target space supersymmetry~\footnote{Preliminary investigations, conducted together with B. Hoare, A. Patella, T. Meier and J. Weber, on the ``supersphere''  setup of Ref.~\cite{Read:2001pz} show that both in the continuum (in dim reg) and on the lattice renormalizability at all loop orders can be proven on the same lines, i.e. exploiting the constraints of the nonlinear symmetry on the possible counterterms. See also the talk of I. Costa at Lattice 2022~\cite{ilaria-lattice}.}. 
%(in terms of functional equations of the effective action). 

This review proceeds as follows. In Section~\ref{sec:general} we sketch general aspects of the GS action in AdS and introduce the model in the continuum (some details of which are in Appendix~\ref{app:continuumdetails}), in Section~\ref{sec:brokensymmetry} we review some of the findings of~\cite{Bianchi:2016cyv,Forini:2016sot,Forini:2017mpu, Bianchi:2019ygz, Bliard:2022kne} and in Section~\ref{sec:symmetrypreserving} we report the more recent analysis of~\cite{Bliard:2022kne}, via which we have a better picture of the challenges underlying this program.
 
\section{The model in the continuum}
\label{sec:general}

The Type IIB string background emerging in the near-horizon limit of D3-branes -- the setup at the heart of the AdS/CFT correspondence -- is the 10-dimensional spacetime 
product of the five-dimensional Anti-de-Sitter space AdS$_5$ and the five-sphere $S^5$, and supported by a self-dual five-form Ramond-Ramond flux~\cite{Maldacena:1997re}.  Because of the presence of the latter and its non-local features when coupled with worldsheet fields, the standard worldsheet supersymmetric formalism of Ramond, Neveu and Schwarz for the string action appears of problematic use. More adequate is the Green-Schwarz (GS) approach, which is supersymmetric only in target space, and whose fundamental fields, both bosonic and fermionic, are worldsheet scalars.  The corresponding TypeIIB GS action in this background can be usefully written %, mimicking the construction in flat space~\cite{Hennauz-mezi}, 
as a Wess-Zumino-Witten-type nonlinear  sigma-model~\footnote{The action is namely the sum of a "kinetic term”, namely a 2d integral quadratic in invariant Cartan 1-forms on the superspace,  and a Wess-Zumino type term, that is an integral of a closed 3-form over a 3d space which has the world-sheet as its boundary. Because the 3-form is closed, in a local coordinate system,  namely choosing a specific parametrization of the Cartan 1-forms in terms of target superspace coordinates,  the action takes the usual 2d sigma-model form.} on the target coset superspace 
$\frac{PSU(2,2|4)}{SO(4,2)\times SO(5)}$~\cite{MT1998},  for a pedagogical review see~\cite{Arytunyovfoundations}.  The global symmetry of the action is the supergroup $PSU(2,2|4)$~\footnote{This is the group of super-isometries (Killing vectors and Killing spinors) of this background as solution of the type IIB 10-d supergravity equations. The stability group  ${SO(4,2)\times SO(5)}$ of the supercoset  follows by writing the AdS$_5$ and $S^5$ themselves as cosets.}, which is the $\mathcal{N}=4$ super-extension of the conformal group in four dimensions SO(4,2). 
The superstring action has two additional local symmetries: a bosonic one, the invariance under reparametrization of the worldsheet or diffeomorphism, and a fermionic one, the  $\kappa$-symmetry~\cite{Siegel:1983hh}, which is crucial to remove unphysical degrees of freedom and ensure space-time supersymmetry of the spectrum. This action is classically integrable, as its equations of motion are equivalent to the flatness condition for a Lax pair and infinitely many non-local conserved charges can be derived~\cite{Bena:2003wd}. About the quantum integrability, for the string worldsheet action this is a solid statement only at low perturbative orders in sigma-model loop expansion, verified through the factorization of the worldsheet S-matrix~\cite{Roiban:2014cia} or because of the fact that 1-loop~\cite{Frolov:2002av} and 2-loop~\cite{Giombi:2009gd} corrections to energies of certain string configurations match the strong-coupling predictions of the $\mathcal{N}=4$ sYM Bethe Ansatz. At finite coupling, integrability remains a conjecture for the whole AdS/CFT system. 
%The Wess-Zumino term guarantees invariance of the full action under that local fermionic symmetry, $\kappa$-symmetry~\cite{Siegel}, which is crucial to remove unphysical degrees of freedom and ensure space-time supersymmetry of the spectrum. 

The $AdS_5\times S^5$ GS sigma-model is expected  to be UV finite. The argument~\cite{MT1998}~\footnote{According to the argument,  any renormalization of both ``kinetic'’ and ``Wess-Zumino’’ terms of the action is excluded, also because of their relation through $\kappa$-symmetry. See for example~\cite{Roiban:2007jf}.} for this is however very formal, and based on the assumption that both global symmetries and the local fermionic $\kappa$-symmetry are actually preserved at the quantum level.  In practice, UV finiteness is not manifest. 
A direct, canonical quantization of the AdS string – namely, deriving the quantum S-matrix from the classically integrable structure -- is beyond current methods, and one proceeds in a semiclassical fashion which begins, usually, by choosing a physical, light-cone gauge-fixing %~\footnote{Alternatively, it is possible to proceed with the conformal gauge~\cite{BOH}.} 
for both diffeomorphism and $\kappa$-symmetry invariance. 
The model is formally power-counting non-renormalizable beyond one loop~\cite{Polyakov:2004br,Roiban:2007jf,Roiban:2007dq} and verifying the cancellation of UV divergences may be non-trivial.  

The original action has no scale --  natural scales are the inverse string tension  $\alpha'$ and the  radius $R$ common to AdS$_5$ and $S^5$, but they combine in a dimensionless combination, the effective string tension  $g=\frac{\sqrt{\lambda}}{4\pi}$, equated via AdS/CFT to the t’Hooft parameter $\lambda$  of the $\mathcal{N}=4$ super Yang-Mills theory.  The bosonic light-cone gauge fixing~$X^+=P_+\tau$~\footnote{For details on the light-cone gauge fixing of interest here, see~\cite{MT2000,Giombi:2009gd}.} where $P_+$ is the (dimensionful) light-cone momentum, introduces however a ``fiducial'' mass scale which below we call $m$.  The fermionic $\kappa$-symmetry light-cone gauge reduces the $32$ fermionic coordinates $\Theta^I_\alpha$ of the two  ($I=1,2$) Type IIB left Majorana-Weyl 10-d spinors to 16 physical Gra\ss mann variables, $\theta_i$ and $\eta_i$ ($i= 1,2,3,4$) below (the action is linear in the first and quartic in the second),  which transform according to the fundamental representations of~$SU(4)$.   
%Another notable feature of the lightcone gauge-fixed sigma-model is that it is not 2d Lorentz-invariant. 

The AdS light-cone gauge-fixed action~\footnote{The light-cone gauge-fixing can be done in two inequivalent ways  -- corresponding to two inequivalent massless geodesics in AdS$_5\times S^5$, one which wraps a big circle on $S^5$ (this is the setup for the famous BMN string solution~\cite{Berenstein:2002jq}), one only running within AdS$_5$, which is what is meant by using ``AdS lightcone''.} is presented in~\cite{Giombi:2009gd}~\footnote{See equations (2.6)-(2.7) there. }. Here we are interested in the action $S_\text{cusp}$ -- equation~\eqref{S_cusp_cont} below -- for the quantum fluctuations  about one of its classical solutions, the one representing an open-string euclidean world surface ending on a closed contour with a light-like cusp on the boundary of AdS. This is because, according to  AdS/CFT~\cite{Maldacena:1998im,Rey:1998ik},  the euclidean  path integral weighted by the action for the quantum fluctuations about this classical solution evaluates the vacuum expectation value of the Wilson loop with the null cusp contour
\begin{equation}\label{Z_cusp}
 \langle W[C_{\rm cusp}]\rangle\equiv  Z_{\rm cusp}= \int [D\delta X] [D\delta\Psi]\, e^{- S_{\rm cusp}[\delta X,\delta\Psi]} = e^{-\Gamma_{\rm eff}}\equiv e^{-\frac{1}{8} f(g)\,m^2\,V_2 }~.                         
\end{equation}
The effective action $\Gamma_{\rm eff}$ to which the path integral is equivalent is then the holographic definition of the cusp anomalous dimension  $f(g)$ governing the renormalization of the Wilson loop~\cite{Belitsky:2006en}.  
The cusp anomaly $f(g)$ can be evaluated perturbatively in gauge theory~\cite{Bern:2006ew} ($g\ll1$) and in sigma-model loop expansion~\cite{Gubser:2002tv,Frolov:2002av,Giombi:2009gd} ($g\gg 1$).  Considering its equivalent definition as leading coefficient in the anomalous dimension of twist operators at large spin~\footnote{The one appearing in~\eqref{Z_cusp} is actually the so-called “scaling function” governing the logarithmic scaling and only equals twice the cusp anomalous dimension of light-like Wilson loops~\cite{Belitsky:2006en}.} and assuming all-order integrability, an integral equation~\cite{BES} can be derived which gives $f(g)$ exactly at each value of the coupling $g$. 
The action $S_{\rm cusp}$ for the bosonic and fermionic fluctuations  $X(t,s),~\Psi(t,s)$ over the cusp minimal surface - with $t,s$ the temporal and spatial coordinate spanning the string worldsheet - has constant coefficients,  hence the worldsheet volume $V_2=\int dt ds$   factorizes out (the normalization factor 1/8 takes also into account the conventions of~\cite{Giombi:2009gd}). The $m$-dependence, in \eqref{Z_cusp} and below, is introduced by dimensional analysis for convenience.

The explicit form for $S_{\rm cusp}$ is~\cite{Giombi:2009gd}
\begin{eqnarray}
\nonumber
S_{\rm cusp}^{\rm cont}=g \int dt ds&& \Bigg\{ 
\left| \partial_t x + \tfrac{m}{2} x \right|^2
+ \tfrac{1}{z^4} \left| \partial_s x - \tfrac{m}{2} x \right|^2
+ \left( \partial_t z^M + \tfrac{m}{2} z^M + \tfrac{i}{z^2} z_N \eta_i \left(\rho^{MN}\right)_{\phantom{i}j}^{i} \eta^j \right)^2
\\ \nonumber &&
+ \tfrac{1}{z^4} \left( \partial_s z^M - \tfrac{m}{2} z^M \right)^2  %\\
+i\, \left( \theta^i \partial_t \theta_i + \eta^i \partial_t \eta_i + \theta_i \partial_t \theta^i + \eta_i \partial_t \eta^i \right) - \tfrac{1}{z^2} \left( \eta^i \eta_i \right)^{2}  \\ \nonumber &&  
+2i \, \Big[
\tfrac{1}{z^3} z^M \eta^i \left( \rho^M \right)_{ij}
\left( \partial_s \theta^j - \tfrac{m}{2} \theta^j - \tfrac{i}{z} \eta^j \left( \partial_s x -\tfrac{m}{2} x \right) \right)
\\ && \qquad
+ \tfrac{1}{z^3} z^M \eta_i \big( {\rho^M}^\dagger \big)^{ij}
\left( \partial_s \theta_j - \tfrac{m}{2} \theta_j + \tfrac{i}{z} \eta_j \left( \partial_s x -\tfrac{m}{2} x \right)^* \right)
\Big]
\,\Bigg\}
\, ,
\label{S_cusp_cont}
\end{eqnarray}
Above, $x$ is a complex bosonic field whose real and imaginary part are the $AdS_5$ (coordinate) fields transverse  to the $AdS_3$ subspace spanned by the classical solution~\eqref{null_cusp_back}. 
$z^M\, (M=1,\cdots, 6)$ are 6 real bosonic fields, $z=\sqrt{z_M z^M}$ is the radial coordinate of the AdS$_5$ space with $z^M/z=u^M $
	identifying points on $S_5$, and together with $x$ they are the bosonic coordinates of the $AdS_5\times S^5$ background in Poincar\'e parametrization remaining after fixing the  light-cone gauge. 
%\colb{In Appendix \ref{app:continuum} we briefly  review the steps leading to the action \eqref{S_cusp}.}  
 %As mentioned above,  the lagrangean above neither contains gauge fields nor actual fermions. Indeed, 
As mentioned above, $\theta_i,\eta_i,\, i=1,2,3,4$ are  4+4 complex anticommuting variables for which  $\theta^i = (\theta_i)^\dagger,$ $\eta^i = (\eta_i)^\dagger$. 
They transform in the fundamental representation of the $SU(4)$ R-symmetry and do not carry (Lorentz) spinor indices.  
The matrices $\rho^{M}_{ij} $ are the off-diagonal
blocks of $SO(6)$ Dirac matrices $\gamma^M$ in the chiral representation
\begin{equation} 
\gamma^M\equiv \begin{pmatrix}
0  & \rho^\dagger_M   \\
 \rho^M   &  0  
\end{pmatrix}
=
\begin{pmatrix}
0  & (\rho^M)^{ij}   \\
(\rho^M)_{ij}   &  0 
\end{pmatrix}
\end{equation}
The two off-diagonal blocks, carrying upper and lower indices respectively, are related by $(\rho^M)^{ij}=-(\rho^M_{ij})^*\equiv(\rho^M_{ji})^*$, so that indeed the block with upper indices, denoted  $(\rho_{M}^{\dagger})^{ij}$, is the conjugate transpose of the block with lower indices.
$(\rho^{MN})_i^{\hphantom{i} j} = (\rho^{[M} \rho^{\dagger N]})_i^{\hphantom{i} j}$ and
$(\rho^{MN})^i_{\hphantom{i} j} = ( \rho^{\dagger [M} \rho^{N]})^i_{\hphantom{i} j}$ are  the
$SO(6)$ generators. Some details are collected in
	Appendix~\ref{app:continuumdetails}.

%The fields $z^M$ are neutral under U(1), $\theta^i$ and $\eta^i$ have opposite
%charges and the charge of $\eta_i$ is half the charge of $x$. 
 The action~\eqref{S_cusp_cont} is invariant under a $U(1)\times SU(4)$ global symmetry acting on the fields as
\begin{gather}
	z^M \to \text{as}(U)^{MN} 
	z^N \ , \quad \theta^i \to U^i_{\phantom{i}j} \theta^j \ , \quad \eta^i \to U^i_{\phantom{i}j} \eta^j \ , \\
	x \to e^{i \alpha} x \ , \quad \theta^i \to e^{i \alpha/2} \theta^i \ , \quad \eta^i \to e^{-i \alpha/2} \eta^j \ ,
\end{gather}
%
%\textcolor{red}{[Or maybe $\theta_i \to U_i^{\phantom{i}j} \theta_j$???]} VF I think it is ok as it is now, with the previous definition of $\theta_i$ it should be understandable 
where $U$ is an element of $SU(4)$ and its representative in the antisymmetric, denoted here as $\text{as}(U)$, is an element of $SO(6)$. The  $U(1)$ is the rotational symmetry in the two $AdS_5$ directions orthogonal  to the classical solution, while the $SU(4)\sim SO(6)$ symmetry originates from the isometries of $S^5$,  unaffected by the (purely) AdS light-cone gauge fixing.  
Notice that the euclidean action~\eqref{S_cusp_cont} is not invariant under worldsheet rotation,  parity ($s \rightarrow -s$), time reversal ($t\rightarrow -t$).\\
The UV finiteness of the model has been verified up to two loops in sigma-model perturbation theory via the evaluation at this order of the cusp anomaly in~\eqref{Z_cusp}~\cite{Giombi:2009gd} and of the ``generalized scaling function'' of~\cite{Giombi:2010zi}, both reproducing the result predicted by integrability~\cite{Basso:2010in}. Also in agreement with the exact result obtained via Bethe Ansatz is the one-loop calculation of the dispersion relation of excitations around the long spinning string 
with one angular momentum in $AdS_5$ and one in $S^5$~\cite{Giombi:2010bj}. The regularization adopted in all these cases is a version of dimensional regularization~\footnote{Given Feynman integrals with components of the loop momenta in the numerators, this prescription consists of performing all manipulations in the numerators in $d= 2$, which has the advantage of simpler tensor integral reductions. Also, one sets to zero power UV divergent massless tadpoles, as in dimensional regularization  $\int d^2p\, (p^2)^n= 0,n\geq 0$. In the examples seen, all logarithmically divergent integrals happen to cancel out in the computation an dthere is no need to pick up an explicit regularization scheme to compute them. See the discussion in~\cite{Roiban:2007dq,abjm2loops}.} introduced in~\cite{Giombi:2009gd,Roiban:2007jf,Roiban:2007dq}.
	
\section{Explorations in the broken symmetry setup}
\label{sec:brokensymmetry}
In this Section we summarise some theoretical and numerical aspects of the simulations addressed in~\cite{Bianchi:2016cyv,Forini:2016sot,Forini:2017mpu, Bianchi:2019ygz}, where -- oriented to simulations -- the model was first linearized (the quartic fermionic interaction being reduced to quadratic, so to be formally integrated out) then discretized with a Wilson-like treatment of the fermionic sector which would break some of the global symmetries of the problem. Simulations were then performed employing a Rational Hybrid Monte Carlo (RHMC) algorithm~\cite{Kennedy:1998cu,Clark:2003na}. 
%tested via a one-loop analysis in lattice perturbation theory. An estimation of the (derivative) of the cusp anomaly of $\mathcal{N}=4$ super Yang-Mills  was also provided, via a measurement of the vacuum expectation value of the relevant action  in terms of simulations performed employing a Rational Hybrid Monte Carlo (RHMC) algorithm.  
%where a fermionic discretization which breaks some of the global symmetries of the problem has been used. 
{While repeating the kind of analysis} in the setup of~\cite{Bliard:2022kne} or similar gauge-fixed ones is first subject to clarifying the question of its finiteness on the lattice, it remains an interesting fact that on these models algorithms and techniques employed in unrelated models with fermionic interactions as well as in lattice QCD could be tested, implemented and shown to improve the analysis. Among these, the procedure of eliminating a complex phase in the fermionic Pfaffian and the attempts to reduce numerical instabilities. 
 %\colb{and general features of the sign problem appearing in the strongly coupled regime}.  
 We do not report on this here,  it is still interesting to recall that in this setup we have  tested~\cite{Forini:2021keh} a stochastic trace estimator algorithm introduced in~\cite{Fitzsimons}, revealing advantages in terms of variance with respect to traditional methods, such e.g. the Gaussian estimator,  and as such of potential benefit in the area of lattice field theory.
 
\bigskip
The AdS light-cone gauge-fixed action is at most quartic in fermions. Having numerical simulations as a goal, the common way to proceed is to linearize the action thus making it quadratic in fermions, then formally integrate them out and re-exponentiate the determinant (here a Pfaffian) obtained in terms of a set of pseudo-fermions so to enter the Boltzmann weight of configurations in the statistical ensemble.

In the case at hand,  one starts with that part of the cusp fluctuation lagrangian Lagrangian~\eqref{S_cusp_cont} which is quartic in fermions  ($n^M=z^M/z$)
 \begin{equation}\label{eq:quarticaction0}
  \mathcal{L}_4=\frac{1}{z^2}\left[- (\eta^2)^2+\left(i\, \eta_i {(\rho^{MN})^i}_j n^N \eta^j\right)^2\right]\,\,,
\end{equation}
and notices  a ``repulsive'' potential, i.e. the plus sign in front of the second term in \eqref{eq:quarticaction0}, that squares an hermitian bilinear $(i\,\eta_i {\rho^{MN}}^i{}_j \eta^j)^\dagger=i\eta_j\,{\rho^{MN}}^j{}_i\,\eta^i$~\cite{Bianchi:2016cyv}. 
This has the effect of introducing a complex phase in the auxiliary, linearized Lagrangian. Indeed, a standard Hubbard-Stratonovich transformation~\cite{McKeown:2013vpa}
\begin{eqnarray}\label{HubbardStratonovich0}
&& \!\!\!\!\!\!\!
\exp \Big\{-g\int dt ds  \Big[-\textstyle{\frac{1}{{z}^{2}}}\left( {\eta}^{i}{\eta}_{i}\right)^{2}  +\Big(\textstyle{\frac{i}{ {z}^{2}}} {z}_{N} {\eta}_{i}{\rho^{MN}}_{\phantom{i}j}^{i} {\eta}^{j}\Big)^{2}\Big]\}\\\nonumber
&& 
\sim\,\int D\phi D\phi^M\,\exp\Big\{-  g\int dt ds\,[\textstyle\frac{1}{2}{\phi}^2+\frac{\sqrt{2}}{z}\phi\,\eta^2 +\frac{1}{2}({\phi}_M)^2-i\,\frac{\sqrt{2}}{z^2}\phi^M {z}_{N} \,\big(i \,{\eta}_{i}{\rho^{MN}}_{\phantom{i}j}^{i} {\eta}^{j}\big)]\Big\}~,
\end{eqnarray}
which introduces $7$ auxiliary fields (one scalar $\phi$ and a SO(6) vector field $\phi^M$),
will generate a non-hermitian term (the last one above) and lead to a  complex-valued Pfaffian~\cite{Bianchi:2016cyv}. To eliminate the complex phase, an alternative procedure -- inspired by the analysis of~\cite{Catterall:2015zua} for the much simpler case of a $SO(4)$ four-fermion interaction  -- was considered in~\cite{Bianchi:2019ygz}, consisting in an algebraic manipulation of the original four-fermion Lagrangian which would essentially change the problematic sign in~\eqref{eq:quarticaction0}. 
After some $\rho$-matrices algebra,  equation~\eqref{eq:quarticaction0} can be written as
\begin{align}\label{newquarticaction}
\mathcal{L}_4=\frac{1}{z^2}\Big(- 4\, (\eta^2)^2+2\big|\eta_i (\rho^N)^{ik} n_N \eta_k\big|^2\Big)\,.
\end{align}
%where the plus sign in front of the second term still prevents a real Pfaffian after the Hubbard-Stratonovich transformation. 
Now one can define the fermionic bilinear ${\S_i}^j\equiv\eta_i \eta^j$ and 
introduce a Hodge-duality-like transformation to obtain its dual  $\tilde{\S}_j{}^i$ as well as its
self- and antiself-dual part~\footnote{Notice that $\tilde{\tilde \S}=\S$ and ${\S^i}_j\equiv ({\S_i}^j)^\dagger={\S_j}^i $.  }
\begin{align}\label{sigmatilde}
\tilde{\S}_j{}^i=n_N n_L(\rho^N)^{ik}(\rho^L)_{jl} {\S_k}^l\,\,,\qquad {\S_{\pm}}=\S \pm \tilde \S\,\qquad \text{such that}~~ \tilde\S_{\pm}=\pm \S_{\pm}\,.
\end{align}
In terms of these, one can write two equivalent forms of the same action, differing by a sign
\begin{align}\label{lagpm}
 \mathcal{L}_4=\frac{1}{z^2}\Tr\left(4 \S\S\mp \S_{\pm} \S_{\pm} \pm 2 \S\S \right)~.
\end{align}
Above, the trace is over $SU(4)$ fundamental indices~\footnote{Notice that $\Tr \tilde \S\tilde \S=\Tr \S\S$.}, and one uses that~$\Tr \S_{\pm}\S_{\pm}=2\,\Tr (\S\S\pm  \S\tilde \S)$.  Notice that there is no ambiguity in the double sign of equation~\eqref{lagpm}: Writing the Lagrangian in terms of the self-dual part of $\S$ requires the minus sign, writing it in terms of the antiself-dual part requires the plus sign. 
It is only in the second case that a complex phase appears, proving that the latter is an artefact of our naive linearization. Choosing the part involving the $\S_+$, one obtains
$\mathcal{L}_4=\frac{1}{z^2}\left(- 6\, (\eta^2)^2 - {\S_{+}}_i^j{\S_{+}}_j^i \right)$ and proceeds with the new Hubbard-Stratonovich transformation
%
 %. In particular we have
 \bal\label{HubbardStratonovich}
&\exp \Big\{-g\int dt ds  \Big[-\textstyle{\frac{1}{z^2}\left(- 6\, (\eta^2)^2 - {\S_{+}}_i^j{\S_{+}}_j^i \right)}\Big]\Big\}\\ 
&\sim ~\int D\phi D\phi_i^j\,\exp\Big\{-  g\int dt ds\,[\textstyle \frac{12}{z} \eta^2 \phi +6\phi^2+\frac{2}{z} {\S_+}^i_j \phi^j_i +\phi^i_j \phi^j_i ]\Big\}~.
\eal
This introduces a total of 17 auxiliary fields ($\phi$ is real, $\phi^i_j$ is a $4\times 4$ complex hermitian matrix with 16 real degrees of freedom), and leads to a Lagrangian now quadratic in fermions
\be\nonumber
%\label{Scuspquadratic}
\!\!\!{\cal L}= \textstyle {| \partial_t {x} + {\frac{m}{2}}{x} |}^2 + \frac{1}{{ z}^4}{\big| \partial_s {x} -\frac{m}{2}{x} |}^2
+ (\partial_t {z}^M + \frac{m}{2}{z}^M )^2 + \frac{1}{{ z}^4} (\partial_s {z}^M -\frac{m}{2}{z}^M)^2
+6\phi^2+\phi^i_j \phi^j_i+\psi^T O_F \psi\
\ee
with $\psi\equiv(\theta^{i},\theta_{i},\eta^{i},\eta_{i})$ and
%%%%%%%%%%%%%%%%%%%%%%%%%%%%%%%%%%%%%%%%%%%%%
\begin{equation} \label{OF}
\!\!\!\!\!\!\!\!\!\! 
O_F =\left(\begin{array}{cccc} 
0 & \mathrm{i}\partial_{t} & -\mathrm{i}\rho^{M}\left(\partial_{s}+\frac{m}{2}\right)\frac{{z}^{M}}{{z}^{3}} & 0\\
\mathrm{i}\partial_{t} & 0 & 0 & -\mathrm{i}\rho_{M}^{\dagger}\left(\partial_{s}+\frac{m}{2}\right)\frac{{z}^{M}}{{z}^{3}}\\
\mathrm{i}\frac{{z}^{M}}{{z}^{3}}\rho^{M}\left(\partial_{s}-\frac{m}{2}\right) & 0 & 2\frac{{z}^{M}}{{z}^{4}}\rho^{M}\left(\partial_{s}{x}-m\frac{{x}}{2}\right) & \mathrm{i}\partial_{t}-A^{T}\\
0 & \mathrm{i}\frac{{z}^{M}}{{z}^{3}}\rho_{M}^{\dagger}\left(\partial_{s}-\frac{m}{2}\right) &\mathrm{i}\partial_{t}+A & -2\frac{{z}^{M}}{{z}^{4}}\rho_{M}^{\dagger}\left(\partial_{s}{x}^\ast-m\frac{{x}}{2}^\ast\right)
\end{array}\right)
\end{equation}
where
\begin{align}\label{A}
A=\textstyle-\frac{6}{z}\phi + \frac{1}{z}\tilde{\phi}+\frac{1}{z^{3}}\rho^\ast_{N}\tilde{\phi}^{T}\rho^{L}z^{N}z^{L}+\mathrm{i}\frac{z^{N}}{z^2}\rho^{MN}\partial_{t}z^{M}, \qquad\qquad
\tilde{\phi}\equiv  \tilde{\phi}_{ij}\equiv \phi^{i}_{j}.
\end{align}
In simpler cases of models with four-fermion interactions~\cite{Catterall:2015zua,Catterall:2016dzf} a similar algebraic manipulation ensures a positive-definite Pfaffian~\footnote{The fermionic operator in~\cite{Catterall:2015zua,Catterall:2016dzf} is \emph{real} and \emph{antisymmetric} -- so that its purely imaginary eigenvalues come in pairs,   and  the symmetries of the model further ensure that all eigenvalues are also doubly degenerate. The latter feature is then the key feature which prevents sign flips as the simulation proceeds, once the Pfaffian is e.g. as the product of eigenvalues with positive imaginary part on the initial configuration of simulations.}.  
In~\cite{Bianchi:2016cyv, Bianchi:2019ygz} a  ($U(1)$-breaking) discretization was adopted,  
 to avoid fermion doublers 
 a Wilson-like term was added in the main diagonal of the fermionic operator. 
The resulting discretized fermionic operator~$\hat O_F$ is antisymmetric and  $\gamma_5$-hermitian'' 
$\hat O_F^\dagger=\Gamma_5\,\hat O_F\, \Gamma_5\,$ (with $\Gamma_5$ a unitary, antihermitian matrix,   $\Gamma_5^\dagger \Gamma_5=\mathbb{1}$, $\Gamma_5^\dagger=-\Gamma_5$) leading to a \emph{real} and \emph{non-negative} $\det \hat O_F$.  A study of the fermionic spectrum~\cite{Bianchi:2019ygz} confirms that the Yukawa-like terms (given in terms of the auxiliary fields resulting from the linearization but also present in the original lagrangian) are responsible for the occurrence of sign flips in the Pfaffian~\footnote{See the discussion in~\cite{Bianchi:2019ygz} about the disposition of eigenvalues in quartets $(\lambda, -\lambda^*, -\lambda, \lambda^*)$ and the role of Yukawa terms for the appearance of purely imaginary or purely real eigenvalues with no degeneracy.}.

The absence of a complex phase is a manifest improvement (which e.g. allows to eliminate systematic errors), but it is not enough to make the Pfaffian positive-definite. Indeed a sign problem has been detected in the simulations of~\cite{Bianchi:2019ygz} for small values of the effective string tension $g$ -- this, in a large tension expansion like the one of sigma-model perturbation theory, identifies the strong coupling regime of the model. 
%
% where working at finite, relatively small values of $N$, number of (square) lattice sites, allowed the the use of exact algorithms as~\cite{wimmer2012algorithm} to evaluate the Pfaffian. 
We have not repeated a similar analysis of the Pfaffian in the symmetry-preserving discretized setup of~\cite{Bliard:2022kne}, see Section~\ref{sec:symmetrypreserving}. However, for this and similar uses of the light-cone gauge fixed action of~\cite{MT2000} (namely, considering the action expanded around string configurations holographically dual to other interesting observables) it is very likely algebraic manipulations of the four-fermion terms as the one described above would be necessary to eliminate complex phases - it is also rather obvious that a sign problem will similarly appear.
%
%It is an interesting question whether the existence of 

Together with the sign problem, for lower values of $g$ (and thus when the string sigma-model is strongly coupled)  simulations in the setup  described here run  into numerical instabilities which can be associated with zero eigenvalues in the fermionic operator and cause the non-convergence of the inverter for the fermionic matrix. This may be cured by a ``twisted-mass'' IR regularization~\cite{Luscher:2008tw}, adding a massive term to the fermionic matrix ($\mu$ being a real parameter)
\be\label{twistedmass}
\tilde O_F=\hat O_F+i\,\mu\,\Gamma_5\,,\qquad\qquad  \tilde O_F {\tilde O_F}^\dagger= \hat O_F {\hat O_F}^\dagger+\mu^2\,\mathbb{1}\,
\ee
so that the $\mu^2\,\mathbb{1}$  term above shifts the eigenvalues of $ \hat O_F {\hat O_F}^\dagger$ apart from zero.  

Concretely, then, simulations are not done with the string worldsheet action in the chosen discretization, but differ due to both the replacement of the Pfaffian by its absolute value \emph{and} the addition of the  ``twisted mass'' in \eqref{twistedmass}.  The sign of the Pfaffian and the low modes of $O_F$ are then taken into account by a standard reweighting procedure, for which the expectation values $\langle \mathcal{O}\rangle$ of observables in the underlying, target  theory are obtained from the  expectation values $\langle O \rangle_\text{m}$ in the theory with the modified, positive-definite  fermionic determinant $( \det\big(\tilde O_F {\tilde O_F}^\dagger)+\mu^2\big)^{\frac{1}{4}}$ as follows
\be\label{reweight}
\langle \mathcal{O}\rangle=\frac{\langle\mathcal{O}\,W\rangle_\text{m}}{\langle W \rangle_\text{m}}\,,
\ee
and the total reweighting factor  $W$  in this  case reads
\be\label{reweight_factors}
W=W_\text{s}\,W_\mu\,,\qquad\qquad W_s=\text{sign}\,\text{Pf}\,\hat O_F\,\qquad\qquad W_\mu=\frac{(\det \hat O_F^\dagger \,\hat O_F)^{\frac{1}{4}}}{\big(\det (\hat O_F^\dagger \,\hat O_F+\mu^2)\big)^\frac{1}{4}}~.
\ee
In the setup of~\cite{Bianchi:2019ygz} -- where two different values for $\mu$ were chosen -- this procedure lead to stability of the simulations in a very large region of the parameter space. We refer the reader to the detailed analysis there included, but report that in the case of numerical values of the ensemble averages for the  two-point functions of bosonic and fermionic fluctuations over the cusp (the configurations being generated by the standard RHMC algorithm) the  \emph{sign}-reweighting seemed \emph{not} to have effect on the measured observables.

\section{Symmetry-preserving discretization and fine-tuning}
\label{sec:symmetrypreserving}

We review here the analysis of~\cite{Bliard:2022kne}, in which for the null cusp effective worldsheet action a new discretization has been worked out, which  is invariant under the full $U(1)\times SU(4)$ group of internal symmetries. 

The proposed action 
\begin{eqnarray}\nonumber
\!\!\!\!\!\!\!\!\!	
S_{\rm cusp}=g
\sum_{s, t} a^2&&\!\!\! \!\!\Bigg\{ \!\!
\left| b_+ \hat\partial_t x + \tfrac{m}{2} x \right|^2
\!\! + \tfrac{1}{z^4}  \left| b_- \hat\partial_s x - \tfrac{m}{2} x \right|^2
\!\!+ \big( b_+ \hat\partial_t z^M + \tfrac{m}{2} z^M + \tfrac{i}{z^2} z^N \eta_i (\rho^{MN})_{\phantom{i}j}^i \eta^j \big)^2
\\ \nonumber && 
+ \tfrac{1}{z^4} \big( \hat\partial_s z^M \hat\partial_s z^M + \tfrac{m^2}{4} z^2 \big)
+ 2 i\, \big( \theta^i \hat\partial_t \theta_i + \eta^i \hat\partial_t \eta_i \big)
- \tfrac{1}{z^2} \left( \eta^i \eta_i \right)^2  \\\nonumber
&& 
+ 2i\, \Big[ \tfrac{1}{z^3} z^M \eta^i \left(\rho^M\right)_{ij}
\big( b_+ \bar\partial_s \theta^j - \tfrac{m}{2} \theta^j - \tfrac{i}{z} \eta^j \big( b_- \hat\partial_s x - \tfrac{m}{2} x \big) \big)
\\\label{S_cusp_latt} && \qquad
+ \tfrac{1}{z^3} z^M \eta_i \big({\rho^M}^\dagger\big)^{ij} \big( b_+ \bar\partial_s \theta_j - \tfrac{m}{2} \theta_j + \tfrac{i}{z} \eta_j \big( b_- \hat\partial_s x^* -\tfrac{m}{2} x^* \big) \!\big)\!
\Big]\Bigg\}\,,
\, 
\end{eqnarray}
where $a$ is the lattice spacing and $b_\pm$ are  two auxiliary parameters whose role will become clear below. 
The action is written, for both the bosonic and the fermionic sector, in terms of the forward and backward discrete derivatives
\begin{equation}
\hat{\partial}_\mu f(\sigma)\equiv\frac{f\left(\sigma +a e_\mu\right)-f\left(\sigma\right)}{a}
\, , \qquad 
\bar{\partial}_\mu f(\sigma)\equiv\frac{f\left(\sigma\right)-f\left(\sigma -a e_\mu\right)}{a}
\, ,
% &=\frac{1}{\mathcal{N}} \sum_{p_{i}} \frac{e^{i p(x+a)}-e^{i p x}}{a} f(p) \\ &=\frac{1}{\mathcal{N}} \sum_{p_{i}} i \hat{p} f(p) e^{i p x}e^{i \frac{p a}{2}}  \quad \hat{p}=\frac{2}{a} \sin \left(\frac{p a}{2}\right). \end{aligned}
\end{equation}
with $e_\mu$ is the unit vector in the direction $\mu=0,1$, and  $\sigma$ is a shorthand notation for $(s,t)$, the worldsheet coordinates. Using forward and backward discrete derivatives
breaks parity and time-reversal,  which is the reason why this is normally avoided for fields satisfying first-order equations of motion - which is usually the case for fermions. As noticed above, however, parity and time-reversal are not symmetries of the action even in the continuum, so that their use is harmless here and - in fact - is the key fact allowing the preservation of the full group of internal symmetries. 

Notice that in the naive continuum limit $a \to 0$, setting $b_\pm= 1$ in~\eqref{S_cusp_latt} gives back the 
 continuum action $S^\text{cont}_{\rm cusp}$ in equation~\eqref{S_cusp_cont}~\footnote{Notice that in the continuum action~\eqref{S_cusp_cont} the term $1/z^4 (z^M \partial_s z^M)$ is actually a vanishing boundary term (this can be seen recalling that $z^M z^M = z^2$). Omitting  it from the discretized version of the action is then possible. This choice, together with the introduction of the $b_\pm$ parameters  in the way specified by~\eqref{S_cusp_latt} is important to get the propagators as below and - with the fine-tuning discussed - is responsible for canceling divergences and reproducing the continuum results for observables.}.  The one loop analysis of~\cite{Bliard:2022kne},  summarized below, shows however that with this choice of the parameters no cancellation of UV divergences occurs. For this to happen, they must be set as in~\eqref{choice}, so that they effectively play the role of fine-tuning parameters. 

 Choosing a flat measure for the fields~\footnote{Notice that this choice is
rather arbitrary, as it is not invariant under reparametrization of the target
AdS$_5 \times S^5$ target space. }, the expectation values of a generic observable $A$  in the lattice discretized theory is then defined by
\begin{equation}
\label{path_int_latt}
\langle A \rangle = \frac{1}{Z_{\text{cusp}}} \int dx dx^* d^6z d^4\theta d^4\theta^\dag d^4\eta d^4\eta^\dag
\, e^{-S_{\text{cusp}}} A
\, ,\qquad\qquad d f \equiv \prod_{s,t} d f(s,t)
\end{equation}
The partition function $Z_{\text{cusp}}$ is as usually fixed by the requirement $\langle 1 \rangle = 1$. 

In fact, for an analysis in lattice perturbation theory one proceeds reparametrizing the bosonic fluctuations as in continuum perturbation theory~\cite{Giombi:2009gd} - see Appendix~\ref{app:continuumdetails}, equations~\eqref{u}-\eqref{exp6} -- so that the 
 corresponding Jacobian determinant  contributes to an 
 effective action  
\begin{gather}
	S_{\text{eff}} = S_{\text{cusp}} - \sum_{s,t} \left\{ 6\phi+ 5\log\left( 1+\frac{y^2}{4} \right) \right\}
	\, ,
	\label{eq:Seff}
\end{gather}
and the path integral~\eqref{path_int_latt} is reformulated as
\begin{equation}
\langle A \rangle = \frac{1}{Z_{\text{eff}}} \int dx dx^* d\phi d^5y d^4\theta d^4\theta^\dag d^4\eta d^4\eta^\dag
\, e^{-S_{\text{eff}}} A
\, .
\end{equation}

The sigma-model perturbative expansion goes in inverse powers of the effective string tension $g$ and it effectively 
splits the action in the sum of its free (quadratic in fluctuations)  and interacting part,  $S_{\text{eff}} = S_0 + S_{\text{int}}$, with 
$S_0$ proportional to $g$ (the $g$-independent quadratic terms coming from
the expansion of the Jacobian determinant belong to $S_{\text{int}}$). 
The 
quadratic action
\begin{eqnarray}
\nonumber
S_0 =
g \,a^2
\sum_{s, t}
\Bigg\{ &&
\left| b_+ \hat\partial_t x + \tfrac{m}{2} x\right|^2
+ \left| b_- \hat\partial_s x - \tfrac{m}{2} x \right|^2
\\ \nonumber &&
+ b_+^2 (\hat\partial_t y^a)^2 + m b_+ y^a \hat\partial_t y^a
+ (\hat\partial_s y^a)^2
\\ \nonumber &&
+ b_+^2 (\hat\partial_t \phi)^2 + m b_+ \phi \hat \partial_t \phi
+ (\hat\partial_s \phi)^2 + m^2 \phi^2
+2 i\left( \theta^i \hat{\partial}_t \theta_i +\eta^i \hat{\partial}_t \eta_i \right)
\\ &&
+2 i \eta^i (\rho^6)_{ij} \left( b_+ \bar{\partial}_s \theta^j - \tfrac{m}{2} \theta^j
\right)
+2 i \eta_i ({\rho^6}^\dagger)^{ij} \left( b_+ \bar{\partial}_s \theta_j - \tfrac{m}{2} \theta_j \right)
\Bigg\}
\, .
\label{S0}
\end{eqnarray}
can be conveniently written in momentum space as
\begin{gather}
	S_0 = g \int_{-\pi/a}^{\pi/a} \frac{d^2p}{(2\pi)^2}
	\left\{ \tilde{\Phi}^t(-p) K_\text{B}(p) \tilde{\Phi}(p) + \tilde{\Psi}^t(-p) K_\text{F}(p) \tilde{\Psi}(p) \right\}
	\, ,
\end{gather}
where  
\begin{eqnarray} 
\Phi&=& (  \text{Re}\,x ,  \text{Im}\,x , y^1 , \dots , y^5, \phi )^t\\
\Psi&=& ( \theta_1 , \dots , \theta_4 , \theta^1 , \dots , \theta^4 , \eta_1 , \dots , \eta_4 , \eta^1 , \dots , \eta^4 )^t
\end{eqnarray}
collect the bosonic and fermionic fields respectively, and $K_\text{B}$, $K_\text{F}$ are the corresponding quadratic operators.  $K_\text{B}(p)$ is a $8 \times 8$ diagonal matrix with non-vanishing components given by
\begin{equation}
K_B^{(n,n)}(p)
=
\begin{cases}
c_+ |\hat{p}_0|^2 + c_- |\hat{p}_1|^2 + \frac{m^2}{2}
\qquad & \text{if } n=1,2 \\
c_+ |\hat{p}_0|^2 + |\hat{p}_1|^2
\qquad & \text{if } n=3,\dots,7\,,\qquad\qquad c_\pm = b_\pm^2 \mp \frac{amb_\pm}{2} \\
c_+ |\hat{p}_0|^2 + |\hat{p}_1|^2 + m^2
\qquad & \text{if } n=8
\end{cases}
\ ,
\label{Eq KB}\\
\end{equation}
where we defined
\begin{gather}
	\hat{p}_\mu = e^{i \frac{a p_\mu}{2}} \frac{2}{a} \sin \frac{a p_\mu}{2}
	\, ,
\end{gather}
and $K_\text{F}(p)$ is the following $16 \times 16$ matrix~\footnote{One uses that $\rho^6 = (\rho^6)^* = - (\rho^6)^t = - {\rho^6}^\dag$, see Appendix~\ref{app:continuumdetails} for details}
\begin{align} 
	\label{Eq KF}
	K_F(p) =  \begin{pmatrix}
		0 &
		- \hat{p}_0^* I_{4 \times 4} &
		- \rho^6 \left( b_+ \hat{p}_1-\frac{im}{2} \right) &
		0
		\\
		- \hat{p}_0 I_{4 \times 4} &
		0 & 
		0 &
		\rho^6 \left( b_+ \hat{p}_1-\frac{im}{2} \right)
		\\
		\rho^6 \left( b_+ \hat{p}_1^*+\frac{im}{2} \right) &
		0 &
		0 &
		- \hat{p}_0^* I_{4 \times 4}
		\\
		0 &
		- \rho^6 \left( b_+ \hat{p}_1^*+\frac{im}{2} \right) &
		- \hat{p}_0 I_{4 \times 4} &
		0
	\end{pmatrix}\,. 
\end{align}
For the two matrices it holds that $K_B^t(p) = K_B(-p)$ and $K_F^t(p) = -K_F(-p)$. Propagators in momentum space are defined by the entries of the inverted matrices up to prefactors~\footnote{To invert $K_F(p)$ one notices that
$K_F(p)^2 = \big( |\hat{p}_0|^2 + c_+ |\hat{p}_1|^2 + \textstyle\frac{m^2}{4} \big) I_{16 \times 16}$. 
} and read explicitly ($\sigma\equiv(s,t)$) %
\begin{eqnarray}\nonumber
&&\!\!\!\!\!\!\!
\textstyle	\sum_{\sigma} a^2 \, e^{-ip \sigma}
	\langle x(\sigma)x^*(0) \rangle_0
	=
	\frac{1}{g}
	\frac{1}{
		c_+ |\hat{p}_0|^2 + c_- |\hat{p}_1|^2 + \frac{m^2}{2}
	}
	\ , \qquad
%	\\
\textstyle	\sum_{\sigma} a^2 \, e^{-ip \sigma}
	\langle y^a(\sigma)y^b(0) \rangle_0
	=
	\frac{1}{2g}
	\frac{\delta^{ab}}{
		c_+ |\hat{p}_0|^2 +  |\hat{p}_1|^2
	}
	\ , \\
&&\!\!\!\!\!\!\!
\textstyle	\sum_{\sigma} a^2 \, e^{-ip \sigma}
	\langle \phi(\sigma)\phi(0) \rangle_0
	=
	\frac{1}{2g}
	\frac{1}{
		c_+ |\hat{p}_0|^2 + |\hat{p}_1|^2 + m^2
	}
	\\\nonumber
&&\!\!\!\!\!\!\!
	\textstyle		\sum_{\sigma} a^2 \, e^{-ip \sigma}
	\langle \theta_i(\sigma) \theta^j(0) \rangle_0
	=
	- \frac{1}{2g}
	\frac{
		\hat{p}_0^* \delta_i^j
	}{
		|\hat{p}_0|^2 + c_+ |\hat{p}_1|^2 + \frac{m^2}{4}
	}
	\ ,
	\qquad
	\sum_{\sigma} a^2 \, e^{-ip \sigma}
	\langle\eta_i(\sigma)\eta^j(0)\rangle_0
	=
	- \frac{1}{2g}
	\frac{
		\hat{p}_0^* \delta_i^j
	}{
		|\hat{p}_0|^2 + c_+ |\hat{p}_1|^2 + \frac{m^2}{4}
	}
%	\ , \\
	% \langle \tilde{\theta}^i(p) \theta_j(0) \rangle_0
	% =
	% \langle\tilde{\eta}^i(p)\eta_j(0)\rangle_0
	% =
	% - \frac{1}{2g}
	% \frac{
	% \hat{p}_0 \delta_j^i
	% }{
	% |\hat{p}_0|^2 + c_+ |\hat{p}_1|^2 + \frac{m^2}{4}
	% }
	% \ , 
	\\ \nonumber
&&\!\!\!\!\!\!\!
\textstyle	\sum_{\sigma} a^2 \, e^{-ip \sigma}
	\langle \theta_i (\sigma) \eta_j(0)\rangle_0
	=
	- \frac{1}{2g}
	\frac{
		\rho^6_{ij} \left( b_+ \hat{p}_1-\frac{im}{2} \right)
	}{
		|\hat{p}_0|^2 + c_+ |\hat{p}_1|^2 + \frac{m^2}{4}
	}
	\ , 
	%\\
	\qquad
	\sum_{\sigma} a^2 \, e^{-ip \sigma}
	\langle \theta^i(\sigma) \eta^j(0)\rangle_0
	=
	- \frac{1}{2g}
	\frac{
		({\rho^6}^\dag)^{ij} \left( b_+ \hat{p}_1-\frac{im}{2} \right)
	}{
		|\hat{p}_0|^2 + c_+ |\hat{p}_1|^2 + \frac{m^2}{4}
	}
	\ ,
\end{eqnarray}
and all other 2-point functions vanish. Notice that the choice $c_\pm = 1$, at which the auxiliary parameters in the discretized action read
\begin{gather}\label{choice}
	\bar{b}_\pm =  \sqrt{ 1 + \left( \frac{am}{4} \right)^2 } \pm \frac{am}{4}
	\ ,
\end{gather}
corresponds to a particularly simple form of the propagators. In fact, it is also the choice which 
reproduces the continuum results for the observables calculated in~\cite{Bliard:2022kne},  see below. 

The interaction Lagrangian is obtained expanding further $S_{\rm eff}$ in powers of the fields, taking into account the explicit expressions of the terms involving the forward derivatives
\begin{eqnarray}
\hat\partial_k z^M(\sigma)
%= &&
%\frac{e^{\phi(x+a e_k)} u^M(x+a e_k) - e^{\phi(x)} u^M(x)}{a}
%\nonumber \\ = &&
%\frac{e^{\phi(x) + a \hat\partial_k\phi(x)} [ u^M(x) + a \hat\partial_k u^M(x) ] - e^{\phi(x)} u^M(x)}{a}
%\nonumber \\ = &&
&=&  e^{\phi(\sigma)} \Big[
\hat\partial_k\phi(\sigma) u^M(\sigma)
+ \hat\partial_k u^M(\sigma)
+ \textstyle\frac{e^{a \hat\partial_k\phi(\sigma)} - 1 - a \hat\partial_k\phi(\sigma)}{a} u^M(\sigma)
\Big]
\ .\\
	\hat\partial_k u^6(\sigma)
	&=& 
	% \frac{1}{a} \left\{ \frac{1-\frac{1}{4}[ y^b(x) + a \hat\partial_k y^b(x) ]^2}{1+\frac{1}{4}[ y^b(x) + a \hat\partial_k y^b(x) ]^2} - \frac{1-\frac{1}{4}y(x)^2}{1+\frac{1}{4}y(x)^2} \right\}
	% = \\
	\frac{
		- 2 y^c(\sigma) \hat\partial_k y^c(\sigma) - a [\hat\partial_k y^c(\sigma)]^2
	}{2 \big[1+\frac{1}{4}[ y^c(\sigma) + a \hat\partial_k y^c(\sigma) ]^2\big]\,\big[ 1+\frac{1}{4}y(\sigma)^2 \big]}
	\ , \\
	\hat\partial_k u^b(\sigma)
	&=& 
	% \frac{1}{a} \left\{
	% \frac{y^b(x) + a \hat\partial_k y^b(x)}{1+\frac{1}{4}[y^b(x)+a \hat\partial_k y^b(x)]^2}
	% - \frac{y^b(x)}{1+\frac{1}{4}y(x)^2}
	% \right\}
	% = \\
	\frac{   
		- 2 y^c(\sigma) \hat\partial_k y^c(\sigma) - a [\hat\partial_k y^c(\sigma)]^2
	}{
		4 \big[ 1+\frac{1}{4}[y^c(\sigma)+a\hat\partial_k y^c(\sigma)]^2\big]\,
		\big[ 1+\frac{1}{4}y(\sigma)^2 \big]
	} \,y^b(\sigma)
\end{eqnarray}
By expanding the exponentials in the first line and denominators in the second and third line in powers of $y$ 
one obtain terms with an arbitrary number of powers of $\hat\partial_k\phi(\sigma)$ and  $\hat\partial_k y^c(\sigma)$, respectively, multiplied by
explicit powers of $a$. The number of derivatives and the number of factors of $a$
are related by dimensional analysis. 

At each order in the perturbative expansion  the interaction Lagrangian density in $\sigma$ is a polynomial of the
fields $\Phi(x)$, $\Psi(x)$, their first derivatives $\hat\partial \Phi(x)$,
$\hat\partial \Psi(x)$, $\bar\partial \Psi(x)$, the lattice spacing $a$, and the
mass $m$. We do not write down explicitly all vertices, but notice that  the possible vertices -- which must have dimension $2$ -- are constrained by dimensional analysis (bosonic 
	fields have mass dimension 0, fermions have dimension 1/2, 
	discrete derivatives as well as $m$ have dimension 1, the lattice spacing
	has dimension -1),   exist only with 0, 2, or 4 fermion fields and 
are proportional to	$m^0$, $m^1$ or $m^2$, and to 	$a^p$ with $p \ge -2$. In particular, terms proportional to $a^{-2}$ are generated by the Jacobian determinant in eq.~\eqref{eq:Seff}.

\bigskip
An analysis of the superficial degree of divergence of a generic Feynman diagram for the lattice-discretized theory -- which can be performed following~\cite{Reisz1988} (see also e.g. \cite{Luscher1988,Capitani2003}) -- shows that it is non-renormalizable by power counting. One can indeed calculate the superficial degree of divergence of the generic one-particle irreducible Feynman diagram $A$ contributing to an amputated $n$-point function in momentum space, which has the general form
\begin{equation}\label{eq:loop_integral}
A
=
\int_{-\frac{\pi}{a}}^{\frac{\pi}{a}} \frac{d^{2} q_{1}}{(2 \pi)^{2}} \cdots \int_{-\frac{\pi}{a}}^{\frac{\pi}{a}} \frac{d^{2} q_{L}}{(2 \pi)^{2}}
W(\hat{p}, \hat{l}; m, a) \prod_{i=1}^I D_i(\hat{l}_i; m, a)
\, ,
\end{equation}
where $q_{i=1,\dots,L}$ are the loop momenta,  $W$ is
the product of all vertices, $p_{i=1,\dots,E}$ denotes the external momenta and $l_{i=1,\dots, I}$ denotes the momentum flowing in the $i$-th internal line, $D_i$ is the propagator associated to the $i$-th internal line. 
The  analysis in~\cite{Bliard:2022kne}~\footnote{See Section 4 there.} leads to the following result for the degree of divergence of $A$ 
\begin{gather}
	\deg A = 2 - \frac{1}{2} E_F - P_m - D_E
	\ ,
	\label{eq:dimA-3}
\end{gather}
where $E_F$ is the number of external fermionic lines, $P_m$ is the total number of $m$ factors and $D_E$ is the total number of discrete derivatives acting on external lines.  The formula shows that the degree of divergence of one-particle irreducible
diagrams cannot be larger than $2$, but a key fact is that $\deg A$ does \emph{not} depend on the  number of external bosonic legs. Therefore  at any loop order the number of divergent diagrams is infinite, implying the need of infinitely many
counterterms at any loop order to cancel the UV divergences. 
As mentioned in Section~\ref{sec:general}, this is not surprising as the same conclusion holds in the continuum~\footnote{From the point of view of the analysis in~\cite{Bliard:2022kne}, in which the total number $P_a$ of factors of $a$ appears at intermediate steps, this can be seen as the fact that  Feynman diagrams appearing in the continuum are the same one as the diagrams with $P_a=0$ on the lattice.}, see~\cite{Polyakov:2004br,Roiban:2007jf,Roiban:2007dq}, and yet -- provided dimensional regularization is adopted (see above) -- non-trivial cancellations of UV divergences happen. This is not the case for the lattice regularization provided here, as a certain amount of fine-tuning is needed in order to reproduce the continuum results at one loop order in the sigma-model expansion. 
%Without extra constraints on the counterterms one would conclude that the theory is non-renormalizable.

\bigskip
A useful way to define the cusp anomaly, free from normalization ambiguities (see discussion in~\cite{Bliard:2022kne}) is through the following derivative 
\begin{gather}
	f(g,m,a) = \frac{4}{m} \frac{\partial}{\partial m} \rho(g,m,a) 
	\label{eq:f_rho}
\end{gather}
of the free energy density in the infinite-volume limit
\begin{gather}
	\rho(g,m,a) = - \lim_{V_2 \to \infty} \frac{1}{V_2} \log
	Z_{\text{cusp}}(g,m,a,V_2) \ ,
\end{gather}
consistently with the definition~\eqref{Z_cusp}. 

At quadratic order the relevant calculation is a Gaussian integral and yields 
\begin{gather}
	\rho(g,m,a)
	=
	g \frac{m^2}{2} - \frac{4}{a^2} \log (2\pi) + \frac{1}{2} \int_{-\pi/a}^{\pi/a} \frac{d^2q}{(2\pi)^2} \log\left[ \frac{\det K_B(q)}{\det K_F(q)} \right]
	+ O(g^{-1})
	\ .
\end{gather}
where
\begin{gather}
	\frac{\det K_B(q)}{\det K_F(q)}
	= 
	\tfrac{
		\left( c_+ |\hat{q}_0|^2 + c_- |\hat{q}_1|^2 + \frac{m^2}{2} \right)^2
		\left( c_+ |\hat{q}_0|^2 + |\hat{q}_1|^2 \right)^5
		\left( c_+ |\hat{q}_0|^2 + |\hat{q}_1|^2 + m^2 \right)
	}{
		\left( |\hat{q}_0|^2 + c_+ |\hat{q}_1|^2 + \frac{m^2}{4} \right)^8
	}
	\ .
\end{gather}
A small-$a$ asymptotic expansion of the relevant integral, whose details can be found in~\cite{Bliard:2022kne}~\footnote{In particular, see Appendix B of~\cite{Bliard:2022kne}. }, leads to 
%using the above asymptotic expansion, with the convention $c_\pm = 1 + am \delta
%c_\pm$, after a lengthy but straightforward calculation, one gets
%
\begin{eqnarray}
\rho(g,m,a)
& = &
g \frac{m^2}{2} 
- \frac{4 \log (2\pi)}{a^2}
+ \frac{m \delta c_-}{2a}
- \frac{3 m^2  \log 2}{8\pi}
- \frac{m^2 \delta c_-^2}{4}
\nonumber \\ &&
+ \frac{m^2 \delta c_- (\delta c_- - 2 \delta c_+)}{4\pi}
+ O(a \log a) + O(g^{-1})
\ ,
\end{eqnarray}
where the convention $c_\pm = 1 + a\,m \,\delta c_\pm$ has been used. For the cusp anomaly one then obtains
\be
	f(g,m,a)
	=
	4 g 
	+ \frac{\delta c_-}{2a m}
	- \frac{3 \log 2}{\pi}
	- 2 \delta c_-^2
	+ \frac{ 2 \delta c_- (\delta c_- - 2 \delta c_+) }{ \pi }
	+ O(a \log a) + O(g^{-1})
	\ .
\ee
Using the naive choice $b_\pm=1$ for the auxiliary parameters, corresponding to $\delta c_\pm
= \mp 1/2$, the cusp anomaly contains a linear divergence. On the other hand,
with the special choice $b_\pm=\bar{b}_\pm$, see~\eqref{choice}, which corresponds to $c_\pm = 1$ and
$\delta c_\pm = 0$, the linear divergence is canceled, and we obtain the same
same result as in dimensional regularization~\cite{Giombi:2009gd,Frolov:2002av}, 
\begin{gather}
	f(g,m,0)
	=
	4 g 
	- \frac{3\log 2}{\pi}
	+ O(g^{-1})
	\ .
\end{gather}

Turning to the calculation of correlators, it is easy to realize that the symmetry-preserving discretization leads to some vanishing one-point functions: $\langle x \rangle = 0$  because of the $U(1)$ symmetry, and $\langle y^a \rangle
= 0$ because of the $SO(5) \subset SO(6) \simeq SU(4)$ symmetry, the rotations for $y^a, \,a=1,\dots,5$.  
%Both in the continuum (in dimensional regularization~\cite{Giombi,Giombi:2010bj,Giombi2010}) and on the lattice, however, 
The fluctuation $\phi$, however, acquires a non-trivial UV-divergent one-point function at one loop.  
%$\phi$ is the only field with a non-vanishing
%one-point function, which has been calculated in dimensional
%regularization~\cite{Giombi,Giombi:2010bj,Giombi2010}.  
In fact, since $\langle\phi\rangle$ appears as a subdiagram in any other $n$-point function, its UV divergence contributes to any physical observable and its inclusion, both in the continuum (in dimensional regularization~\cite{Giombi:2009gd,Giombi:2010bj,Giombi:2010fa}) and on the lattice, is important for the cancellation of divergences - as we see below. 

There are two classes of vertices contributing to the one-point function of
$\phi$: single-field vertices coming from the measure $S_\phi \sim g^0$, which produces a tree-level diagram, and three-field vertices coming from the action $S_{\phi \bullet \bullet}\sim g$, which produces a one-loop  (in the $x$ fields and in the fermions) diagram - for explicit expressions and more details see~\cite{Bliard:2022kne}. The result  (because of the mismatch in the power of $g$) is a contribution to the same order in $g$, yielding
\begin{eqnarray}
\langle \phi \rangle
&=&
\frac{3}{g m^2 a^2}
+ \frac{2}{g m^2}
\int_{-\pi/a}^{\pi/a} \frac{d^2q}{(2\pi)^2}
\frac{
	c_- |\hat{q}_1|^2 + \tfrac{m^2}{4}
}{
	c_+ |\hat{q}_0|^2 + c_- |\hat{q}_1|^2 + \frac{m^2}{2}
}
\nonumber \\ &&
- \frac{1}{2gm^2}
\int_{-\pi/a}^{\pi/a} \frac{d^2q}{(2\pi)^2}
\frac{
	c_+ |\hat{q}_0|^2 - |\hat{q}_1|^2
}{
	c_+ |\hat{q}_0|^2 + |\hat{q}_1|^2 + m^2
}
- \frac{5}{2gm^2}
\int_{-\pi/a}^{\pi/a} \frac{d^2q}{(2\pi)^2}
\frac{
	c_+ |\hat{q}_0|^2 - |\hat{q}_1|^2
}{
	c_+ |\hat{q}_0|^2 + |\hat{q}_1|^2
}
\nonumber \\ &&
- \frac{8}{gm^2}
\int_{-\pi/a}^{\pi/a} \frac{d^2q}{(2\pi)^2}
\frac{
	c_+ |\hat{q}_1|^2 + \frac{m^2}{4}
}{
	|\hat{q}_0|^2 + c_+ |\hat{q}_1|^2 + \frac{m^2}{4}
}
+ O(g^{-2})
\label{eq:phi-1}
\ .
\end{eqnarray}

With the special choice of the auxiliary parameters $b_\pm=\bar{b}_\pm$ in~\eqref{choice} ($c_\pm = 1$), 
one can use the
symmetry $p_0 \leftrightarrow p_1$ of the integrals to obtain the logarithmically divergent integral
\begin{eqnarray}
\langle \phi \rangle
&=&
- \frac{1}{g}
\int_{-\pi/a}^{\pi/a} \frac{d^2q}{(2\pi)^2}
\frac{
	1
}{
	|\hat{q}|^2 + \frac{m^2}{4}
} + O(g^{-2})
\nonumber \\ &=&
\frac{1}{g} \left\{
\frac{1}{4\pi} \log \frac{(am)^2}{4}
+ \frac{1}{4\pi}
- I_0^{(0,0)} + O(a \log a)
\right\} + O(g^{-2})\,,
\label{eq:phi-2}
\end{eqnarray}
where the numerical constant $I_0^{(0,0)} \simeq 0.355$ is defined through an integral~\footnote{Explicitely, it is 
$
I_0^{(0,0)} = I_0(1,1) =
\frac{1-\gamma}{4\pi}
+ \int_0^1 ds \, [K(s)]^2
+ \int_1^\infty ds \, \left\{ [K(s)]^2 - \frac{1}{4\pi s} \right\}\simeq 0.355
$
with the definition
$
K(s)
= 
\int_{-\pi}^{\pi} \frac{d z}{2\pi}
e^{- 4 s \sin^2 \frac{z}{2} }
=
\frac{1}{\sqrt{4\pi s}} + O(s^{-2})
\ . %\label{eq:app:int1:K}
$
}, see Appendix B in~\cite{Bliard:2022kne}. Interestingly, measure, fermion-loop and $x$-loop contributions develop quadratic divergences, whose cancellation is highly non-trivial.
In the more general case $c_\pm = 1 + a\,m\,\delta c_\pm$ where $\delta c_\pm =
O(a^0)$, the asymptotic expansions of the integrals lead to s
\be
\!\!\!\!\!\!
\langle \phi \rangle
=
\frac{1}{g} \bigg\{
\frac{- 8 \delta c_+ + \delta c_-}{\pi a}
+ \frac{1}{4\pi} \log \frac{(am)^2}{4}
+ \frac{1}{4\pi}
- I_0^{(0,0)} 
+ \frac{8 \delta c_+^2 - \delta c_-^2}{2\pi}
+ O(a \log a) \bigg\} + O(g^{-2})
\label{eq:phi-3}
\ , 
\ee
so that the naive choice $b_\pm = 1$, or $\delta c_\pm
= \mp 1/2$,  yields indeed a linear divergence for $\langle \phi \rangle$:
\begin{gather}
\langle \phi \rangle
=
\frac{1}{g} \left\{ \frac{
	9
}{2\pi a}
+ O(\log a) \right\} + O(g^{-2})
\label{eq:phi-4}
\ .
\end{gather}

\begin{figure}
	\centering
	\includegraphics[scale=0.6]{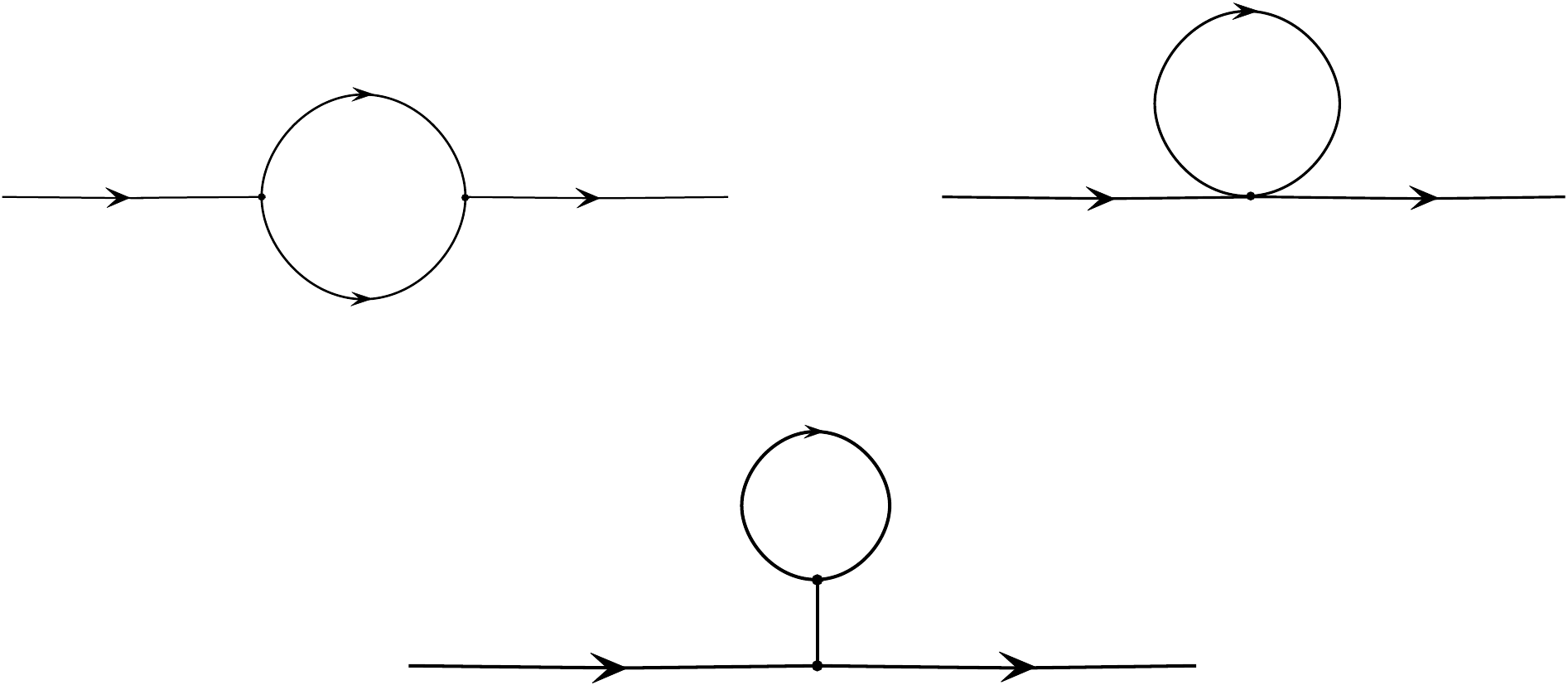}
	\caption{Topologies of diagrams contributing to the two point function at 1-loop.}
	\label{fig:conn_two_p}
\end{figure}

\bigskip
The last calculation we sketch evaluates the two-point function of the bosonic excitation $x$ at
one loop, which we calculate to extract the corresponding dispersion relation. 
The two classes of vertices contributing are three-field vertices $S_{x x^* \bullet}$
and four-field vertices	$S_{x x^* \bullet \bullet}$, with resulting  Feynman diagrams of  three different topologies
illustrated in Fig.~\ref{fig:conn_two_p}, where the tadpole contribution is proportional to $\langle \phi \rangle$. Explicit expressions and more details are, again, in~\cite{Bliard:2022kne}.

The two-point function can be expressed, on general grounds, in the  form
\begin{equation}\!\!\!\!\! 
\langle \tilde{x}(p)x^*(0) \rangle
=
\frac{1}{g}
\left\{
c_+ |\hat{p}_0|^2 + c_- |\hat{p}_1|^2 + \frac{m^2}{2}
+ \frac{1}{g} \left( c_- |\hat{p}_1|^2 + \frac{m^2}{4} \right) \Pi_a(p) + O(g^{-2})
\right\}^{-1}
\ . \label{eq:2pt-x}
\end{equation}
where factor $\left(c_- |\hat{p}_1|^2 + \tfrac{m^2}{4} \right)$ comes from the Fourier transform of the combination
$\left(b_- \hat{\partial}_s x - \tfrac{m}{2} x \right)$, or its complex
conjugate, appearing in all interaction vertices with $x$. Above, $\Pi_a(p)$ can be represented in terms of amputated
Feynman diagrams, and reads explicitly  %
\begin{eqnarray}
\Pi_a(p)
&=&
- 4 g \langle \phi \rangle
+ 4 \int_{-\pi/a}^{\pi/a} \frac{d^2q}{(2\pi)^2}
\frac{1}{c_+ |\hat{q}_0|^2 + |\hat{q}_1|^2 + m^2}
\nonumber \\ &&
- 8 \int_{-\pi/a}^{\pi/a} \frac{d^2q}{(2\pi)^2}
\frac{c_- |\hat{q}_1|^2 + \frac{m^2}{4}}{c_+ |\hat{q}_0|^2 + c_- |\hat{q}_1|^2 + \frac{m^2}{2}}
\frac{1}{
	c_+ |\widehat{p+q}_0|^2 + |\widehat{p+q}_1|^2 + m^2
}
\nonumber \\ &&
- 8 \int_{-\pi/a}^{\pi/a} \frac{d^2q}{(2\pi)^2}
\frac{\hat{q}_0}{|\hat{q}_0|^2 + c_+ |\hat{q}_1|^2 + \frac{m^2}{4}}
\frac{\widehat{p+q}_0^*}{
	|\widehat{p+q}_0|^2 + c_+ |\widehat{p+q}_1|^2 + \frac{m^2}{4}
}
\ .
\end{eqnarray}
As seen above, the  $\langle \phi \rangle$ terms contains in general a linear divergence, while all remaining integrals are logarithmically divergent. 
Using $c_\pm = 1$, which can be done up to terms that vanish in the $a \to 0$ limit, one obtains the simpler expression
\begin{eqnarray}
\Pi_a(p)
&=&
- 4 g \langle \phi \rangle
+ 4 \int_{-\pi/a}^{\pi/a} \frac{d^2q}{(2\pi)^2}
\frac{1}{|\hat{q}|^2 + m^2}
- 8 \int_{-\pi/a}^{\pi/a} \frac{d^2q}{(2\pi)^2}
\frac{|\hat{q}_1|^2 + \frac{m^2}{4}}{|\hat{q}|^2 + \frac{m^2}{2}}
\frac{1}{
	|\widehat{p+q}|^2 + m^2
}
\nonumber \\ &&
- 8 \int_{-\pi/a}^{\pi/a} \frac{d^2q}{(2\pi)^2}
\frac{\hat{q}_0}{|\hat{q}|^2 + \frac{m^2}{4}}
\frac{\widehat{p+q}_0^*}{
	|\widehat{p+q}|^2 + \frac{m^2}{4}
}
+ O(a \log a)
\ .
\end{eqnarray}
The leading divergence of these integrals does not
depend on the external momentum, exactly as it happens in the continuum. Then the subtracted $\Delta
\Pi_a(p) = \Pi_a(p) - \Pi_a(0)$ has the finite $a \to 0$ limit
\begin{eqnarray}
\Delta \Pi_0(p)
&=&
- 8 \int_{-\infty}^\infty \frac{d^2q}{(2\pi)^2}
\frac{q_1^2 + \frac{m^2}{4}}{q^2 + \frac{m^2}{2}}
\left\{ \frac{1}{(p+q)^2 + m^2} - \frac{1}{q^2 + m^2} \right\}
\nonumber \\ \label{Deltapi0} &&
- 8 \int_{-\infty}^\infty \frac{d^2q}{(2\pi)^2}
\frac{q_0}{|\hat{q}|^2 + \frac{m^2}{4}}
\left\{
\frac{p_0+q_0}{(p+q)^2 + \frac{m^2}{4}}
- \frac{q_0}{q^2 + \frac{m^2}{4}}
\right\}
+ O(a \log a)
\ ,
\end{eqnarray}
and all the divergences are contained in
\begin{gather}
	\Pi_a(0)
	=
	% - 4 g \langle \phi \rangle
	% - 4 \int_{-\pi/a}^{\pi/a} \frac{d^2q}{(2\pi)^2}
	% \frac{
	% |\hat{q}|^2
	% }{
	% \left( |\hat{q}|^2 + \frac{m^2}{4} \right)^2
	% }
	% + O(a \log a)
	% = \\ =
	% - 4 g \langle \phi \rangle
	% - 4 \int_{-\pi/a}^{\pi/a} \frac{d^2q}{(2\pi)^2}
	% \frac{
	% 1
	% }{
	% |\hat{q}|^2 + \frac{m^2}{4}
	% }
	% + m^2 \int_{-\infty}^\infty \frac{d^2q}{(2\pi)^2}
	% \frac{
	% 1
	% }{
	% \left( q^2 + \frac{m^2}{4} \right)^2
	% }
	% + O(a \log a)
	% = \\ =
	- 4 g \langle \phi \rangle
	- 4 \int_{-\pi/a}^{\pi/a} \frac{d^2q}{(2\pi)^2}
	\frac{
		1
	}{
		|\hat{q}|^2 + \frac{m^2}{4}
	}
	+ \frac{1}{\pi}
	+ O(a \log a)
	\ .
\end{gather}
%
%where we have used the symmetry of the integrals under $p_0 \leftrightarrow p_1$
%exchange to simplify them.

Using now the result ~\eqref{eq:phi-2} for the tadpole contribution and the choice $c_\pm = 1$,  it is immediate to see that
all divergences cancel and $\Pi_0(0) = 1/\pi$. The two-point function is finite
in the continuum limit and 
\begin{gather}
	\lim_{a \to 0} \langle \tilde{x}(p)x^*(0) \rangle
	=
	\frac{1}{g}
	\left\{
	p^2 + \frac{m^2}{2}
	+ \frac{1}{g} \left( p_1^2 + \frac{m^2}{4} \right) \Pi_0(p) + O(g^{-2})
	\right\}^{-1}\,.
\end{gather}
The two-point function has poles at $p_0 = \pm i E(p_1)$ for every value of
$p_1$, where $E(p_1)$ is the energy of a single excitation propagating on the worldsheet with momentum $p_1$. In
the continuum limit one obtains~\cite{Bliard:2022kne}
\begin{eqnarray} 
E(p_1)^2 &=& p_1^2 + \frac{m^2}{2}
+ \frac{1}{g} \left( p_1^2 + \frac{m^2}{4} \right) \Pi_0\left(\sqrt{p_1^2 + \frac{m^2}{2}},p_1\right) + O(g^{-2})
\nonumber \\ &=&
p_1^2 + \frac{m^2}{2}
- \frac{1}{gm^2} \left( p_1^2 + \frac{m^2}{4} \right)^2
+ O(g^{-2})
\ , 
\label{eq:disprelation}
\end{eqnarray}
%
%where we have used the on-shell value of $\Pi_0$~\eqref{Pi0}. The obtained
%dispersion relation 
reproducing the dispersion relation of \cite{Giombi:2010bj}~\footnote{To compare with~\cite{Giombi:2010bj}, a  redefinition of the worldsheet coordinates is necessary, equivalent to rescaling the fluctuations square masses with a factor of $4$.}. 

However in the general case $c_\pm = 1 + (am) \delta c_\pm$ where $\delta c_\pm
= O(a^0)$, both the integral $\Pi_a(0)$ and the dispersion relation $E(p_1)$ inherit from  $\langle
\phi \rangle$ the linear divergence. For example,  the naive choice $b_\pm = 1$ leads to 
\be
	E(p_1)^2 = p_1^2 + \frac{m^2}{2}
	+ \frac{1}{g} \left( p_1^2 + \frac{m^2}{4} \right)
	\left[ - \frac{18}{\pi a} + O(\log a) \right]
	+ O(g^{-2})
	\ .
\ee
As discussed in~\cite{Bliard:2022kne}, with the 
naive choice of auxiliary parameters $b_\pm = 1$, the divergence
in the dispersion relation cannot be eliminated by renormalizing the remaining
available parameters, i.e. the effective string tension $g$ and the massive parameter $m$. This means that the choice $b_\pm = 1$
is not stable under renormalization. On the other hand, allowing the
coefficients $b_\pm$ to be renormalized together with $m$ and $g$,  the
divergences in the dispersion relation are eliminated for example choosing
\begin{gather}
	b_+ = 1 + \frac{1}{g_R} \frac{ \frac{am_R}{8} }{ 2 + \frac{am_R}{2} } \left( \Pi_a(0) - \frac{1}{\pi} \right) 
	\ , \\
	b_- = 1 - \frac{1}{g_R} \frac{ 1 + \frac{5am_R}{8} }{ 2 + \frac{am_R}{2} } \left( \Pi_a(0) - \frac{1}{\pi} \right) 
	\ , \\
	m^2 = m_R^2 \left[ 1 + \frac{1}{2g_R} \left( \Pi_a(0) - \frac{1}{\pi} \right) \right]
	\ , \\
	g = g_R \left[ 1 + O(g^{-1}) \right]
	\ .
\end{gather}
The continuum dispersion relation has then the same form as eq.~\eqref{eq:disprelation} but with the renormalized $m_R$ replacing $m$. 
Similarly, a one-loop renormalization of the coupling constant could be chosen so that the
cusp anomaly is finite. 
Such observation do not imply that the chosen lattice theory is renormalizable, something which we do not know.  
If, however,  the lattice theory is renormalizable, it appears not sufficient to renormalize $m$ and $g$. One also needs to introduce extra
coefficients in the action and either fine-tune their tree-level value, or renormalize them.

We conclude summarizing some features of the divergences emerging in the one loop analysis above. About \emph{quadratic divergences}, the cancel at one loop in the one-point function
	of $\phi$ and in the two-point function of $x$ (they are instead subtracted by
	hand in the cusp anomaly). The expectation is that such cancellations will always happen in any reasonable
	discretization of the action.
About \emph{linear divergences}, they generally do not cancel in all considered observables. They arise from the special choice of forward and
	backward discrete derivatives and are therefore very specific of the
	lattice discretization. As mentioned, they are cured via the introduction of two extra parameters $b_\pm$ in the action, to be either fine-tuned at tree level or
	renormalized at one-loop.
  Once the linear divergences are removed this way, 
	the \emph{logarithmic} ones cancel in the cusp anomaly and in the two-point
	function of $x$ (while they survive in the one-point function of $\phi$, which is what also happens in the continuum). 
	Then the continuum limit of the observables we calculated is the same as in dimensional
	regularization.

\section*{Acknowledgements}
	
	We are grateful to Agostino Patella for collaboration on the recent~\cite{Bliard:2022kne}, and to all coauthors - in particular Bj\"orn Leder - of the papers on the subject reviewed here: Lorenzo Bianchi, Marco Stefano Bianchi, Philipp Töpfer, Edoardo Vescovi, Nils Wauschkuhn.  
	The research of GB is funded from the European Union's Horizon 2020 research and innovation programme under the Marie Sklodowska-Curie ITN grant No 813942.    
	The research of IC, and partially of GB, is funded by the Deutsche Forschungsgemeinschaft (DFG, German Research Foundation) - Projektnummer 417533893/GRK2575 "Rethinking Quantum Field Theory". 
	The research of VF is supported by the STFC grant ST/S005803/1, the European ITN grant No 813942 and from the Kolleg Mathematik Physik Berlin.

	%%%%%%%%%%%%%%%%%%%%%%%%%%%%%%%%%%%%%
	%%%%%%%%%%%%%%%%%%%%%%%%%%%%%%%%%%%%%

\appendix	
\section{The model in the continuum: details}
\label{app:continuumdetails}	
	
In this Appendix we shortly review the steps leading to the action~\eqref{S_cusp_cont}.
 
One starts with the $AdS_5\times S^5$ metric in Poincar\'e coordinates  (here the common radius of both $AdS_5$ and $S^5$ is set to $1$)
\begin{eqnarray}\nonumber
&&ds^2=z^{-2}\,(dx^m\,dx_m+dz^M\,dz^M)=z^{-2}(dx^m\,dx_m+dz^2)+du^M du^M\\\label{adsmetric}
&&x^m x_m=x^+ x^-+x^* x\,,\qquad x^\pm=x^3\pm x^0\,,\qquad x=x^1+i x^2\,,\\\nonumber
&&z^M=z\,u^M\,, \qquad u^M\,u^M=1\,\qquad z=(z^M z^M)^{\frac{1}{2}}~,
\end{eqnarray}
where $x^\pm$ are the light-cone coordinates, $x^m=(x^0,x^1,x^2, x^3)$ parametrize the 
four-dimensional  boundary of $AdS_5$ and $z\equiv e^{\phi}$ is the radial coordinate.

The  AdS light-cone gauge~\cite{MT2000,MTT2000} is defined by fixing the local symmetries of the superstring action, bosonic diffeomorphisms and $\kappa$-symmetry
 as follows~\footnote{As in the standard conformal gauge,  the choice $x^+ = p^+ \tau$ is allowed by residual
diffeomorphisms after the choice \eqref{bosgauge}.}
\begin{eqnarray}  \label{bosgauge}
&&  \sqrt{-g} g^{\a\b} = {\rm diag}(-z^2, z^{-2})\ , \qquad \qquad x^+ = p^+ \tau \ ,\\
&&\Gamma^+ \theta^I=0\,.
\end{eqnarray}
The  resulting \adss   superstring  action  can be written as % ($z^M=z \, u^M$)
\begin{eqnarray}
\label{s}
S &=& \frac{1}{2} T \int d \tau \int 
 d \sigma \; \mathcal{L}\ , \quad  \quad  \quad T =
\frac{R^2}{2 \pi \alpha'} = \frac {\sqrt{\lambda}}{2 \pi} \ , \\
\mathcal{L} &=& \dot{x}^* \dot{x} + (\dot z^M  + \mathrm{i}  p^+ z^{-2} z^N 
\eta_i {\rho^{MN}}^i{}_j \eta^j)^2  + \mathrm{i} p^+ (\theta^i \dot{\theta}_i +
       \eta^i\dot{\eta}_i +\theta_{i}\dot{\theta}^{i}+\eta_{i}\dot{\eta}^{i}
)+\no \\
&&
  \quad  - (p^+)^2 z^{-2} (\eta^2)^2 - z^{-4} ( x'^*x'  + {z'}^M {z'}^M) 
  \no \\
&&
  \quad - 2 \Big[\ p^+ 
       z^{-3}\eta^i \rho_{ij}^M z^M (\theta'^j - \mathrm{i}
       z^{-1} \eta^j  x') +  p^+ 
       z^{-3}\eta^i (\rho^\dagger_M)^{ij} z^M (\theta'^j + \mathrm{i}
       z^{-1} \eta^j  x'^*\Big]\; \label{la}\\
       &\equiv&\dot{x}^* \dot{x} + (\dot z^M  + \mathrm{i}  p^+ z^{-2} z^N 
\eta_i {\rho^{MN}}^i{}_j \eta^j)^2  + \mathrm{i} p^+ (\theta^i \dot{\theta}_i +
       \eta^i\dot{\eta}_i - h.c.) - (p^+)^2 z^{-2} (\eta^2)^2 \no \\
&&
  \quad - z^{-4} ( x'^*x'  + {z'}^M {z'}^M) - 2 \Big[\ p^+ 
       z^{-3}\eta^i \rho_{ij}^M z^M (\theta'^j - \mathrm{i}
       z^{-1} \eta^j  x') + h.c.\Big]\;.  \label{la}
\end{eqnarray}
%The action \eqref{s}  is \emph{real}. 
%~\footnote{For the terms in the round and square brackets one should take into account, respectively, that
%\begin{eqnarray}
%&&\Big( \mathrm{i}  \eta_i {\rho^{MN}}^i{}_j \eta^j\Big)^\dagger=-\mathrm{i}\,(\eta^j)^\dagger({\rho^{MN}}^i{}_j)^*(\eta_i)^\dagger
%=-\mathrm{i}\,\eta_j\,{\rho^{MN}}_i{}^j\,\eta^i=+\mathrm{i}\,\eta_j\,{\rho^{MN}}^j{}_i\,\eta^i\equiv   \mathrm{i}  \eta_i {\rho^{MN}}^i{}_j \eta^j
%\\
%&&\Big(\eta^i(\rho^M)_{ij}\theta'^j\Big)^\dagger=(\th'^j)^\dagger(\rho^M_{ij})^*(\eta^i)^\dagger=-\th'_j(\rho^M)^{ij}\eta_i=\eta_i(\rho^M)^{ij}\theta'_j~.
%\end{eqnarray}
 %}, and 
 % enters the partition function as $Z=e^{i\,S}$. 
 Wick-rotating $\tau \to  -\mathrm{i} \tau, \ p^+ \to  \mathrm{i} p^+$, and setting $p^+=1$, one gets $Z=e^{-S_E}$, where $S_E = \frac{1}{2} T \int d \tau d \sigma \; \mathcal{L}_E$  and 
%%%%%%%%%%%%%%%%%%%%%%%%%%%%%%%%%%%%%%%%%%%%%
\begin{flalign}\nonumber
\mathcal{L}_{E} & =\dot{x}^{*}\dot{x}+ \big(\dot{z}^{M}+i\,z^{-2}z_{N}\eta_{i}(\rho^{MN})^i_{\hphantom{i} j}\eta^{j}\big)^{2}+i\big(\theta^{i}\dot{\theta}_{i}+\eta^{i}\dot{\eta}_{i}-\text{h.c.}\big)-z^{-2}\left(\eta^2\right)^{2}\\\label{LCADS_euc}
&+z^{-4} (x^{'*}x^{'}+z^{'M}z^{'M})
 +2i\Big[z^{-3}z^{M}\eta^{i}{\rho^{M}}_{ij}\big(\theta^{'j}-i\,z^{-1}\eta^{j}x^{'}\big)+\text{h.c.}\Big]
 \end{flalign}
%\footnote{The l.c. gauge-fixed \emph{euclidean} action is not real because of the term in square brackets, since $(i[...+\text{h.c.}])^\dagger=-i (...+\text{h.c.})$. 

%%%%%%%
\noindent
The null cusp background
\be
\label{null_cusp_back}
x^{+}=\tau \qquad \qquad x^{-}=-\frac{1}{2 \sigma} \qquad \qquad x=x^{*}=0 \qquad \qquad z=\sqrt{\frac{\tau}{\sigma}}\,,\qquad\qquad  \tau,\sigma>0\,,
\ee
is the classical solution of the string action that describes a Euclidean open string surface ending on a lightlike Wilson cusp in the AdS boundary at $z=0$ \cite{Giombi:2009gd}. This string vacuum is actually degenerate as any $SO(6)$ transformation on $z^M$ leaves the last condition above unaltered. The fluctuation spectrum of this solution can be easily found by fixing a direction, say $u^M=(0,\,0,\,0,\,0,\,0,\,1)$, and defining the fluctuation fields
\begin{eqnarray}
&& z=\sqrt{\frac{\tau}{\sigma}}\ {\tilde z} \ , \ \ \ \ \ \ \ \ 
{\tilde z} = e^{\tilde \phi}= 1 + \tilde \phi  +\dots~,\ \ \  
 z^M=\sqrt{\frac{\tau}{\sigma}}\ {\tilde z}^M \ , \ \ \ \ 
{\tilde z}^M = e^{\tilde \phi} \tilde u^M  \label{u}\\
&&
{\tilde u}{}^{a}=  \frac{y^{a}}{1+\frac{1}{4}y^2}~, \ \ \ \ 
{\tilde u}{}^{6} =  \frac{1-\frac{1}{4}y^2}{1+\frac{1}{4}y^2}  \ , \ \ \ \ \ \ \ \ \
~~~~ y^2\equiv \sum_{a=1}^5 (y^a)^2\ , \ \ \ \ \ a=1,...,5 \ , \label{exp6}\\
&&
x = \sqrt{\frac{\tau}{\sigma}} \ {\tilde x}
~,~~~~~~
\theta=\frac{1}{\sqrt{\sigma}}{\tilde\theta}
~,~~~~~~
\eta=\frac{1}{\sqrt{\sigma}}{\tilde\eta} \,. 
\end{eqnarray}
The further redefinition of the worldsheet coordinates
\begin{gather}
t=\log\tau \qquad \qquad s=\log\sigma\,
\end{gather}
which absorb powers of $\tau,\, \sigma$ so that the resulting fluctuation Lagrangian has constant coefficients, leads to 
the euclidean action~\eqref{S_cusp_cont}. 

%%%%%%%%%%%%%%%%%%%%%%%%%%%%%%%%%
%\bigskip
 
%%%%%%%%%%%%%%%%%%%%%%%%%%%%%%%%%%%%%%%%%%%%%

%For completeness, we  also report here the expression for the complex-conjugate of the quadratic fermionic operator obtained via the Hubbard-Stratonovich transformation \eqref{HubbardStratonovich}
%\begin{flalign}
%\!\!\!\!\!\!\!
% O_F^\dagger & =\left(\begin{array}{cccc}
%0 & i\partial_{t} & \mathrm{i}\rho^\dagger_{M}\left(\partial_{s}+\frac{1}{2}\right)\frac{{z}^{M}}{{z}^{3}} & 0\\
%\mathrm{i}\partial_{t} & 0 & 0 &\mathrm{i}\rho^{M}\left(\partial_{s}+\frac{1}{2}\right)\frac{{z}^{M}}{{z}^{3}}\\
%-\mathrm{i}\frac{{z}^{M}}{{z}^{3}}\rho^\dagger_{M}\left(\partial_{s}-\frac{1}{2}\right) & 0 & 2\frac{{z}^{M}}{{z}^{4}}\rho \dagger_{M}\left(\partial_{s}{x}^*-\frac{{x}^*}{2}\right) & i\partial_{t}+A\\
%0 &- \mathrm{i}\frac{{z}^{M}}{{z}^{3}}\rho^{M}\left(\partial_{s}-\frac{1}{2}\right) &\mathrm{i}\partial_{t}-A^\dagger & -2\frac{{z}^{M}}{{z}^{4}}\rho^{M}\left(\partial_{s}{x}-\frac{{x}}{2}\right)
%\end{array}\right)
%\end{flalign}
%%where the operator $A$ is defined  in \eqref{Aoperator}.  
%%%%%%%%%%%%%%%%%%%%%%%%%%%%%%%%%%%%%%%%%%%%%

In the action \eqref{S_cusp_latt} and of course also in~\eqref{S_cusp_cont}, the matrices $\rho^M$ appear, which are off-diagonal blocks of the six-dimensional Dirac matrices
in chiral representation
\be 
\gamma^M\equiv \begin{pmatrix}
	0  & {\rho^M}^\dagger   \\
	\rho^M   &  0 
\end{pmatrix}
=
\begin{pmatrix}
	0  & (\rho^M)^{ij}   \\
	(\rho^M)_{ij}   &  0 
\end{pmatrix}
\ee
\begin{align}
\rho_{ij}^M &=- \rho_{ji}^M\,, &
({\rho^M}^\dag)^{il}\rho_{lj}^N + ({\rho^N}^\dag)^{il}\rho_{lj}^M
&=2\delta^{MN}\delta_j^i \,.
%=(\rho^M{}^\dagger)^{ij}\ ,~~~~~~~~ 
\end{align}
The two off-diagonal blocks, carrying upper and lower indices respectively, are related by $(\rho^M)^{ij}=-(\rho^M_{ij})^*\equiv(\rho^M_{ji})^*$, so that the block with upper indices, $({\rho^M}^{\dagger})^{ij}$, is the conjugate transpose of the block with lower indices.
A possible explicit  representation is
\begin{equation}
\begin{aligned}
\rho^1_{ij}&=\left(\begin{matrix}0&1&0&0\\-1&0&0&0\\0&0&0&1\\0&0&-1&0
\end{matrix}\right)\,,&
\rho^2_{ij}&=\left(\begin{matrix}0&\mathrm{i}&0&0\\-\mathrm{i}&0&0&0\\0&0&0&-\mathrm{i}\\0&0&\mathrm{i}&0
\end{matrix}\right)\,,&
\rho^3_{ij}&=\left(\begin{matrix}0&0&0&1\\0&0&1&0\\0&-1&0&0\\-1&0&0&0
\end{matrix}\right)\,, \\
\rho^4_{ij}&=\left(\begin{matrix}0&0&0&-\mathrm{i}\\0&0&\mathrm{i}&0\\0&-\mathrm{i}&0&0\\ \mathrm{i}&0&0&0
\end{matrix}\right)\,,&
\rho^5_{ij}&=\left(\begin{matrix}0&0&\mathrm{i}&0\\0&0&0&\mathrm{i}\\-\mathrm{i}&0&0&0\\0&-\mathrm{i}&0&0
\end{matrix}\right)\,,&
\rho^6_{ij}&=\left(\begin{matrix}0&0&1&0\\0&0&0&-1\\-1&0&0&0\\0&1&0&0
\end{matrix}\right)\,.
\end{aligned}
\label{Eq:rho6}
\end{equation}
The $SO(6)$ generators are built out of the $\rho$-matrices via
\begin{equation}
\rho^{MN}{}^i{}_{ j} \equiv\frac{1}{2}[  ({\rho^M}^\dag)^{il}\rho_{lj}^N
- ({\rho^N}^\dag)^{il}\rho_{lj}^M ].
\end{equation}

For the $SO(6)$ generators built out of the $\rho^{M}_{ij} $ of $SO(6)$ Dirac matrices it holds
\begin{equation}\label{rhomatrices}
\begin{split}
  (\rho^{MN})^i_{\hphantom{i}j}&=\frac{1}{2}({\rho^M}^{i\ell}\,\rho^N_{\ell j} -{\rho^N}^{i\ell}\,\rho^M_{\ell j} )=\frac{1}{2}(\rho^M_{i\ell}\,{\rho^N}^{\ell j} -\rho^N_{i\ell}\,{\rho^M}^{\ell j} )^*\equiv\left( (\rho^{MN})_i^{\hphantom{i}j}\right)^*\\  (\rho^{MN})^i_{\hphantom{i}j}&=-(\rho^{MN})_j^{\hphantom{j}i} \,\qquad\qquad  (\rho^{MN})_i^{\hphantom{i}j}=- (\rho^{MN})^j_{\hphantom{j}i}\,,
  \end{split}
\end{equation}
where in the last equation we used that $\frac{1}{2}({\rho^M}^{i\ell}\,\rho^N_{\ell j} -{\rho^N}^{i\ell}\,\rho^M_{\ell j} )=-\frac{1}{2}(\rho^M_{j\ell}\,{\rho^N}^{\ell i} -\rho^N_{j\ell}\,{\rho^M}^{\ell i} )$.
%Then
%\begin{eqnarray}
%\Big(\eta^i (\rho^{MN})_i^{\hphantom{i}j}\theta_j\Big)^\dagger&=&\frac{1}{2}(\theta_j)^\dagger\,[\rho^M_{i\ell}{\rho^N}^{\ell j}-{\rho^N}^{i\ell}\rho^M_{\ell j}]^*(\eta^i)^\dagger=\theta^j(\rho^{MN})^{i}_{\hphantom{i}j}\eta_i\\\nonumber
%&=&-\eta_i \,(\rho^{MN})^{i}_{\hphantom{i}j}\,\theta^j~.
%\ee
Useful flipping rules are
\begin{eqnarray}
\eta\,\rho^M\,\theta&=&\eta^i\,\rho^M_{ij}\,\theta^j=-\theta^j\,\rho^M_{ij}\,\eta^i=\theta^j\,\rho^M_{ji}\,\eta^i\equiv \theta^i\,\rho^M_{ij}\,\eta^j=\theta\,\rho^M\,\eta\\
\eta^\dagger\rho^\dagger_M\,\theta^\dagger&=&\eta_i\,{\rho^M}^{ij}\,\theta_j=-\theta_j\,{\rho^M}^{ij}\,\eta_i=\theta_j\,{\rho^M}^{ji}\,\eta_i\equiv \theta_i\,{\rho^M}^{ij}\,\eta_j=\theta^\dagger \rho^\dagger_M\,\eta^\dagger\\
\eta_i\,(\rho^{MN})^i_{\hphantom{i}j}\,\theta^j&=&-\theta^j\,(\rho^{MN})^i_{\hphantom{i}j}\,\eta_i=\theta^j\,(\rho^{MN})_j^{\hphantom{j}i}\,\eta_i\equiv\theta^i\,(\rho^{MN})_i^{\hphantom{i}j}\,\eta_j~.
\end{eqnarray}

\bigskip

\noindent Data Availability Statement: No Data associated in the manuscript.

	\bibliographystyle{nb}
	\bibliography{Ref_strings_lattice}
	
\end{document}